\documentclass{emulateapj}
\shorttitle{Two LBV Eruptions and their Progenitors}
\shortauthors{Smith et al.}
\begin{document}

\title{Discovery of Precursor LBV Outbursts in Two Recent Optical
  Transients: \\ The Fitfully Variable Missing Links UGC~2773-OT and
  SN~2009\lowercase{ip}}

\author{Nathan Smith\altaffilmark{1,2}, Adam Miller\altaffilmark{1},
  Weidong Li\altaffilmark{1}, Alexei V.\ Filippenko\altaffilmark{1},
  Jeffrey M.\ Silverman\altaffilmark{1}, \\ Andrew W.\
  Howard\altaffilmark{1}, Peter Nugent\altaffilmark{3}, Geoffrey W.\
  Marcy\altaffilmark{1}, Joshua S. Bloom\altaffilmark{1}, Andrea M.\
  Ghez\altaffilmark{4}, \\ Jessica Lu\altaffilmark{4}, Sylvana
  Yelda\altaffilmark{4}, Rebecca A.\ Bernstein\altaffilmark{5}, \&
  Janet E.\ Colucci\altaffilmark{5}}

\altaffiltext{1}{Department of Astronomy, University of California,
  Berkeley, CA 94720-3411.} 
\altaffiltext{2}{email: nathans@astro.berkeley.edu.}
\altaffiltext{3}{Lawrence Berkeley National Laboratory, 1 Cyclotron
  Rd., Berkeley, CA 94720.}
\altaffiltext{4}{Division of Astronomy and Astrophysics, University of
  California, Los Angeles, CA 90095-1547.} 
\altaffiltext{5}{Department of Astronomy and Astrophysics, 1156 High
  Street, UCO/Lick Observatory, University of California, Santa Cruz,
  CA 95064.}

\begin{abstract}

  We present progenitor-star detections, light curves, and optical
  spectra of supernova (SN) 2009ip and the 2009 optical transient in
  UGC~2773 (U2773-OT), which were not genuine supernovae.  Precursor
  variability in the decade before outburst indicates that both of the
  progenitor stars were luminous blue variables (LBVs).  Their
  pre-outburst light curves resemble the S~Doradus phases that
  preceded giant eruptions of the prototypical LBVs $\eta$~Carinae and
  SN~1954J (V12 in NGC~2403), with intermediate progenitor
  luminosities.  {\it Hubble Space Telescope} detections a decade
  before discovery indicate that the SN~2009ip and U2773-OT
  progenitors were supergiants with likely initial masses of 50--80
  M$_{\odot}$ and $\ga$20 M$_{\odot}$, respectively.  Both outbursts
  had spectra befitting known LBVs, although in different physical
  states.  SN~2009ip exhibited a hot LBV spectrum with characteristic
  speeds of 550 km s$^{-1}$, plus evidence for faster material up to
  5000 km s$^{-1}$, resembling the slow Homunculus and fast blast wave
  of $\eta$ Carinae. In contrast, U2773-OT shows a forest of narrow
  absorption and emission lines comparable to that of S~Dor in its
  cool state, plus [Ca~{\sc ii}] emission and an infrared excess
  indicative of dust, similar to SN~2008S and the 2008 optical
  transient in NGC~300 (N300-OT).  The [Ca~{\sc ii}] emission is
  probably tied to a dusty pre-outburst environment, and is not a
  distinguishing property of the outburst mechanism.  The LBV nature
  of SN~2009ip and U2773-OT may provide a critical link between
  historical LBV eruptions, while U2773-OT may provide a link between
  LBVs and the unusual dust-obscured transients SN~2008S and N300-OT.
  Future searches will uncover more examples of precursor LBV
  variability of this kind, providing key clues that may help unravel
  the instability driving LBV eruptions in massive stars.

\end{abstract}


%%%%%%%%%%%%%%%%%%%%%%%%% FIGURE 1 - finder charts %%%%%%%%%%%%%%%%%%%%%%%%%
\begin{figure*}
\epsscale{0.96}
%\plotone{prog.eps}
\plotone{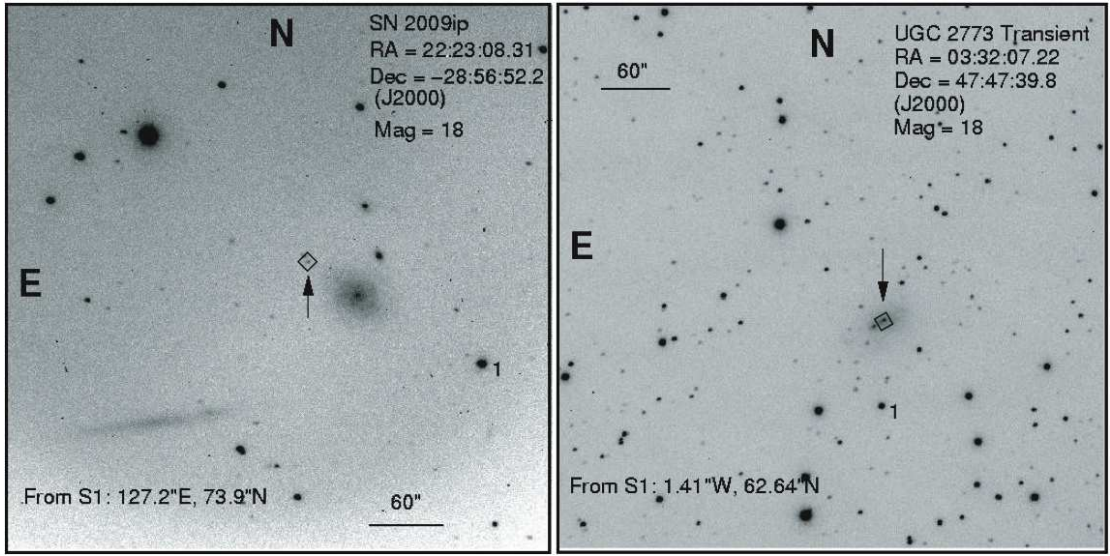}
\caption{Finder charts showing the positions of SN~2009ip and U2773-OT
  with respect to their host galaxies.  The small squares indicate the
  fields of view for the {\it HST}/WFPC2 F606W images in
  Figure~\ref{fig:prog}.  These are unfiltered images obtained with
  the Katzman Automatic Imaging Telescope (KAIT) at Lick Observatory.}
\label{fig:finder}
\end{figure*}
%%%%%%%%%%%%%%%%%%%%%%%%%%%%%%%%%%%%%%%%%%%%%%%%%%%%%%%%%%%%%%%%%%%%%%

%%%%%%%%%%%%%%%%%%%%%%%%% FIGURE 2 - progenitors %%%%%%%%%%%%%%%%%%%%%%%%%
\begin{figure*}
\epsscale{0.98}
%\plotone{prog.eps}
\plotone{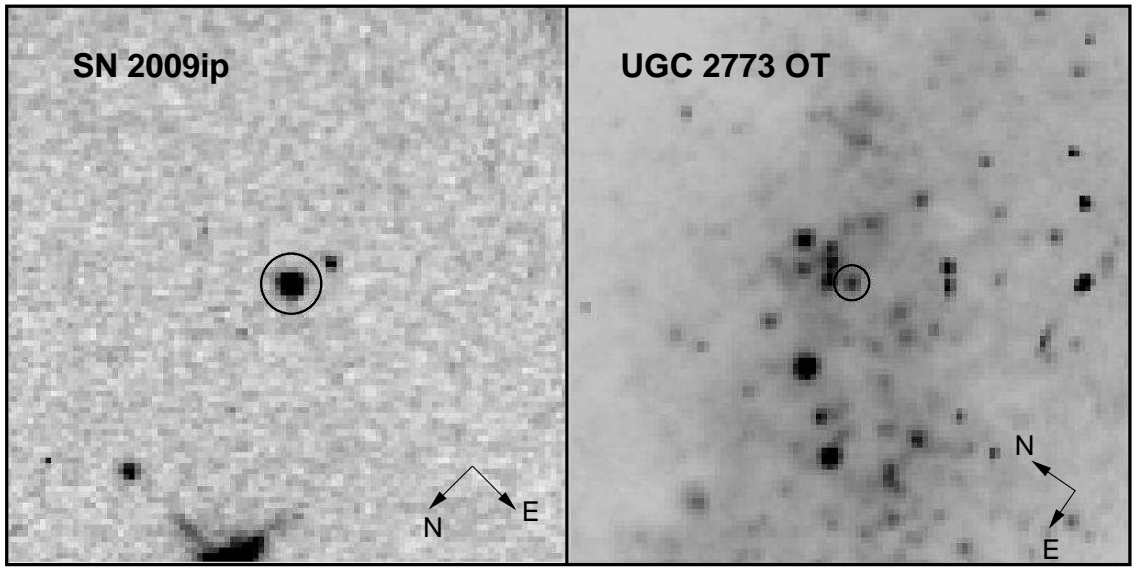}
\caption{The sites of SN~2009ip and U2773-OT in the {\it HST}/WFPC2
  F606W images.  Each stamp is $10\arcsec \times 10\arcsec$. The
  location of each transient as derived from high-resolution
  ground-based images is marked with a circle having a radius that is 15
  times the 1$\sigma$ uncertainty radius of the astrometric solution. A
  good candidate progenitor is identified for both transients.}
\label{fig:prog}
\end{figure*}
%%%%%%%%%%%%%%%%%%%%%%%%%%%%%%%%%%%%%%%%%%%%%%%%%%%%%%%%%%%%%%%%%%%%%%

%%%%%%%%%%%%%%%%%%%%%%%%%%%%%%%%%%%%%%%%%%%%%%%%%%%%%%%%%%%%%%%%%%%%%%%%%%
\section{INTRODUCTION}

Among the objects discovered in the course of hunting for supernovae
(SNe) are transient sources that are fainter and have slower expansion
speeds than most core-collapse SNe.  Following the historical examples
(see Humphreys et al.\ 1999, and references therein) of the 19th
century eruption of $\eta$~Carinae, the 17th century eruption of
P~Cygni, SN~1954J (Variable 12 in NGC~2403), and SN~1961V, these
transients have usually been associated with nonterminal eruptions of
luminous blue variables (LBVs), rather than final explosions that mark
the deaths of massive stars.  LBVs represent a highly unstable, rapid
mass-loss phase in the late evolution of massive stars (see Smith \&
Owocki 2006; Smith et al.\ 2004; Humphreys \& Davidson 1994), which is
still poorly understood.  P~Cygni and $\eta$~Car remain as luminous
well-studied blue supergiants, and the survivor of SN~1954J has been
identified as a luminous dust-enshrouded star (Smith et al.\ 2001; Van
Dyk et al.\ 2005).  SN~1961V is a more controversial case (Chu et al.\
2004), but there are indications of a surviving star as well (Goodrich
et al.\ 1989; Filippenko et al.\ 1995; Van Dyk et al.\ 2002).  Because
the massive stars are thought to survive the events, recent examples
of these $\eta$~Car analogs such as SN~1997bs (Van Dyk et al.\ 2002)
have earned the label ``SN impostors.''  For most recent extragalactic
examples, however, available evidence that the star has survived
remains inconclusive.

Interest in and interpretation of these transients was stirred by the
surprising discovery that two recent events, SN~2008S and the 2008
optical transient in NGC~300 (hereafter N300-OT), both had relatively
low-luminosity, dust-enshrouded progenitor stars (Prieto et al.\ 2008;
Prieto 2008).  Although the outburst properties resembled those of
known LBV eruptions, interpretation of their dusty progenitors (Prieto
et al.\ 2008; Thompson et al.\ 2009) implied initial masses below the
usually recognized initial-mass range for LBVs extending down to
$\sim$20~M$_{\odot}$ (Smith et al.\ 2004).  This fueled a range of
speculation that these transients might be similar eruptive phenomena
extending to somewhat lower mass and cooler stars (Smith et al.\
2009; Berger et al.\ 2009a; Bond et al.\ 2009), the eruptive birth of
a white dwarf and planetary nebula in stars with initial masses below
8 M$_{\odot}$ (Thompson et al.\ 2009), electron-capture SNe (ecSNe) in
extreme asymptotic giant branch (EAGB) stars around 8--10 M$_{\odot}$
(Thompson et al.\ 2009; Botticella et al.\ 2009), faint core-collapse
SNe (see Pastorello et al.\ 2007 in regard to previous events), or
mergers/mass-transfer events related to some other recent transients
(e.g., Kulkarni et al.\ 2007; Kashi et al.\ 2009).  Interpretation of
these objects remains controversial and puzzling.

Gogarten et al.\ (2009) found a likely initial mass of 12--25
M$_{\odot}$ for the N300-OT progenitor based on the star-formation
history of its local neighborhood, challenging the ecSN or white-dwarf
birth hypotheses.  Deriving a likely initial mass around 10--12
M$_{\odot}$ from the luminosity (e.g., Prieto et al.\ 2008) depends
upon the assumption that the progenitor was a cool EAGB star at the
uppermost tip of its AGB.  It remains possible, however, that the
progenitors of SN~2008S and N300-OT were relatively blue stars that
were heavily obscured by circumstellar dust (Prieto et al.\ 2008;
Smith et al.\ 2009; Berger et al.\ 2009a; Bond et al.\ 2009).  In
that case, the initial mass implied by the progenitor's infrared (IR)
luminosity would be closer to 15--20 M$_{\odot}$ for N300-OT, and
would therefore be in better agreement with the findings of Gogarten
et al.\ (2009) than an 8--10 M$_{\odot}$ EAGB star.  The somewhat
lower luminosity of the progenitor of SN~2008S would imply 12--15
M$_{\odot}$ under the same assumption (Smith et al.\ 2009).

In this paper we discuss another pair of newly discovered transients
with identified progenitors, the 2009 optical transient in UGC~2773
(hereafter U2773-OT) and SN~2009ip (see Fig.~\ref{fig:finder}).
U2773-OT occurred in the dwarf irregular galaxy UGC~2773, and was
discovered (Boles 2009) on 2009 Aug.\ 18.08 (UT dates are used
throughout this paper).  SN~2009ip was discovered (Maza et al.\ 2009)
on 2009 Aug.\ 26.11 in the Sb galaxy NGC~7259.  Both objects had
discovery absolute magnitudes fainter than $-$14~mag, and they had
spectra with narrow H emission lines (Berger et al.\ 2009b; Berger \&
Foley 2009).  These transient outbursts, still currently ongoing,
share properties in common with giant LBV eruptions as well as
SN~2008S and N300-OT.  In these new cases, however, the progenitors
are not as heavily obscured by dust.  The progenitors are detected at
optical wavelengths, and they exhibit pre-eruption variability that is
matched by classic LBVs such as V12 and $\eta$~Car.  We reported the
preliminary pre-outburst detections of SN~2009ip in Miller et al.\
(2009), while our photometry of U2773-OT is documented here for the
first time.  Based on their LBV-like pre-outburst variability, the two
new transients help bridge the gap between classical LBV eruptions and
objects similar to SN~2008S and N300-OT.  Other than the historical
cases of SN~1954J and SN~1961V, this is the first discovery of
extended 5--10~yr pre-eruption variability before an extragalactic
LBV-like transient.

%%%%%%%%%%%%%%%%%%%%%%%%%%%%%%%%%%%%%%%%%%%%%%
\section{OBSERVATIONS}

\subsection{Pinpointing the Progenitors}

Both SN~2009ip and U2773-OT had observations taken $\sim$10~yr prior to
discovery with the {\it Hubble Space Telescope}/Wide Field Planetary
Camera 2 ({\it HST}/WFPC2), which we retrieved from the archive and
analyzed.  NGC~7259 (SN~2009ip) was observed in the F606W filter on
1999 Jun.\ 29, and UGC~2773 was observed in the F606W and F814W
filters on 1999 Aug.\ 14.

To pinpoint the precise location of the two transients' progenitors in
the {\it HST} images, we obtained high-resolution ground-based images
for comparison. On 2009 Sep.\ 9, we observed SN~2009ip in the
$K^{\prime}$ band with the Near-Infrared Camera 2 (NIRC2) using the
laser guide star (LGS) adaptive optics (AO) system (Wizinowich et al.\
2006) on the 10-m Keck~II telescope. Three mosaic pointings, each with
three exposures of $4 \times 15$~s, were combined to yield a final
stacked image with 9~min total exposure time, a pixel scale of
0$\farcs$04 pixel$^{-1}$, and a field of view of $41\arcsec \times
41\arcsec$.  For U2773-OT, we took a 20~s guider image with the
high-resolution echelle spectrometer (HIRES; Vogt et al.  1994) on the
10-m Keck~I telescope on 2009 Sep.\ 9.  The image has a pixel scale of
0$\farcs$3 pixel$^{-1}$ and a field of view of $43\arcsec \times
58\arcsec$.

To perform astrometric solutions between the ground-based and {\it
  HST} images, we adopted the technique detailed by Li et al.\ (2007)
using stars present in both the ground-based and {\it HST} images. Due
to the small field of view of the ground-based images, we are only
able to measure the positions for five stars in each field.  The
astrometric solutions using IRAF/GEOMAP yield a precision of 36 mas
and 21 mas for SN 2009ip and U2773-OT, respectively\footnote{Although
  the ground-based AO image of SN~2009ip has much higher resolution
  than the U2773-OT image, the astrometric solution is less precise
  because of the faintness of stars in the image.}. The positions of
the transients are then mapped onto the {\it HST} images.  Due to the
small available number of stars, we only use second-order polynominals
in the astrometic transformations. We also conducted an additional
error analysis by taking out one star and leaving four with which to
perform the solution, and repeating this for all stars. For both
transients, the average position from the five separate measurements
and the associated uncertainty is fully consistent with the position
measured from the solution using all five stars together, so our
solutions appear to be stable despite the small number of stars
involved.

Figure~\ref{fig:prog} shows a $10\arcsec \times 10\arcsec$ region of
the sites for the transients in the {\it HST}/WFPC2 images. A
candidate progenitor is detected for both transients within 1$\sigma$
precision of the astrometric solution.  We therefore confirm the
candidate progenitors marked in Figure~\ref{fig:prog}, first proposed
for SN~2009ip by our group (Miller et al.\ 2009) and for U2773-OT by
Berger \& Foley (2009).  The {\it HST} photometry for the progenitors
is measured with HSTphot (Dolphin 2000a, 2000b) and listed in Tables 1
and 2.  HSTphot also reports that both objects have a stellar
point-spread function (PSF): they belong to type ``1" (good star),
with a sharpness well within the range of $\pm$ 0.3 as suggested by
Leonard et al.  (2008).  The coordinates of the two transients, as
measured directly from the World Coordinate System (WCS) in the WFPC2
FITS images, are $\alpha_{\rm J2000}$ = 3$^{\rm h}$32$^{\rm
  m}$07$\fs$22, $\delta_{\rm J2000}$ =
47$\arcdeg$47$\arcmin$39$\farcs$4 for U2773-OT and $\alpha_{\rm
  J2000}$ = 22$^{\rm h}$23$^{\rm m}$08$\fs$20, $\delta_{\rm J2000}$ =
$-$28$\arcdeg$56$\arcmin$52$\farcs$6 for SN~2009ip.

%%%% TABLE 1 - Photometry
\begin{deluxetable}{lcccl}
\tablewidth{0pc}\tighten
\tablecaption{Photometry of SN~2009\lowercase{ip}}
\tablehead{
\colhead{MJD} &\colhead{filter} &\colhead{mag} &\colhead{$1\sigma$} &\colhead{data source}  }
\startdata
51358.50  &F606W       &21.8      &0.2   &HST \\
53195.47  &$R$         &$>$21.17  &...   &DS \\
53226.33  &$R$         &$>$21.43  &...   &DS \\
53233.25  &$R$         &$>$20.97  &...   &DS \\
53251.23  &$R$         &$>$21.01  &...   &DS \\
53259.21  &$R$         &$>$20.65  &...   &DS \\
53541.44  &$R$         &20.61     &0.14  &DS \\
53554.46  &$R$         &20.37     &0.16  &DS \\
54305.39  &$R$         &20.97     &0.22  &DS \\
54323.35  &$R$         &20.92     &0.25  &DS \\
54701.33  &$R$         &$>$21.51  &...   &DS \\
54702.33  &$R$         &$>$21.33  &...   &DS \\
55043.50  &unfiltered  &18.5      &0.4   &CBET 1928 \\
55069.61  &unfiltered  &17.9      &0.3   &CBET 1928 \\
55071.75  &unfiltered  &17.0      &0.3   &CBET 1928 \\
55073.50  &$R$         &18.2      &0.1   &KAIT \\
55085.10  &unfiltered  &20.2      &0.2   &guider \\
55096.50  &unfiltered  &18.3      &0.2   &guider \\
55097.50  &$R$         &18.3      &0.2   &KAIT \\
\enddata
%\tablecomments{}
%\tablenotetext{a}{}
\end{deluxetable}

%%%% TABLE 2 - Photometry
\begin{deluxetable}{lcccl}
\tablewidth{0pc}\tighten
\tablecaption{Photometry of U2773-OT}
\tablehead{
\colhead{MJD} &\colhead{filter} &\colhead{mag} &\colhead{$1\sigma$}  &\colhead{data source}}
\startdata
51404.1&F606W&22.83&0.03&HST \\
51404.1&F814W&22.22&0.05&HST \\
51853.5&unfiltered&$>$20.80&...&KAIT (stack)\\
52225.5&unfiltered&$>$21.10&...&KAIT (stack)\\
52589.5&unfiltered&$>$21.00&...&KAIT (stack)\\
52939.5&unfiltered&$>$20.60&...&KAIT (stack)\\
53683.0&unfiltered&19.91&0.42&KAIT (stack)\\
54042.5&unfiltered&19.16&0.07&KAIT (stack)\\
54394.9&unfiltered&18.93&0.10&KAIT (stack)\\
54772.1&unfiltered&18.73&0.06&KAIT (stack)\\
55051.5&unfiltered&17.96&0.10&KAIT\\
55061.5&unfiltered&17.70&0.19&KAIT\\
55072.5&unfiltered&17.91&0.06&KAIT\\
55077.5&unfiltered&17.76&0.05&KAIT\\
55079.5&unfiltered&17.76&0.03&KAIT\\
55081.5&unfiltered&17.82&0.04&KAIT\\
55082.5&unfiltered&17.76&0.03&KAIT\\
55083.5&unfiltered&17.79&0.03&KAIT\\
55084.5&unfiltered&17.61&0.12&KAIT\\
55090.5&unfiltered&17.74&0.03&KAIT\\
55091.5&unfiltered&17.73&0.03&KAIT\\
55092.5&unfiltered&17.75&0.03&KAIT\\
55094.5&unfiltered&17.75&0.03&KAIT\\
55095.5&unfiltered&17.72&0.03&KAIT\\
55096.5&unfiltered&17.68&0.03&KAIT\\
55097.5&unfiltered&17.71&0.03&KAIT\\
55098.5&unfiltered&17.67&0.03&KAIT\\
55099.5&unfiltered&17.67&0.03&KAIT\\
55100.5&unfiltered&17.74&0.03&KAIT\\
55102.5&unfiltered&17.71&0.03&KAIT\\
55105.5&unfiltered&17.67&0.04&KAIT\\
55110.5&unfiltered&17.69&0.05&KAIT\\
55113.5&unfiltered&17.71&0.03&KAIT\\
55120.5&unfiltered&17.65&0.03&KAIT\\
55123.5&unfiltered&17.59&0.03&KAIT\\
55126.5&unfiltered&17.62&0.03&KAIT\\
55129.5&unfiltered&17.57&0.03&KAIT\\
55132.5&unfiltered&17.55&0.03&KAIT\\
55133.5&unfiltered&17.57&0.03&KAIT\\
55136.5&unfiltered&17.67&0.04&KAIT\\
55149.5&unfiltered&17.52&0.03&KAIT\\
55152.5&unfiltered&17.57&0.03&KAIT\\
55155.5&unfiltered&17.65&0.03&KAIT\\
55158.5&unfiltered&17.60&0.03&KAIT\\
55161.5&unfiltered&17.72&0.03&KAIT\\
55164.5&unfiltered&17.56&0.03&KAIT\\
55169.5&unfiltered&17.66&0.03&KAIT\\
55173.5&unfiltered&17.60&0.05&KAIT\\
\enddata
%\tablecomments{}
%\tablenotetext{a}{}
\end{deluxetable}

%%%% TABLE 3 - IR photometry of U2773-OT
\begin{deluxetable}{lccc}
\tablewidth{0pc}\tighten
\tablecaption{PAIRITEL Photometry of U2773-OT}
\tablehead{
\colhead{MJD} &\colhead{$J$ (mag)} &\colhead{$H$ (mag)} &\colhead{$K$ (mag)}}
\startdata
55082.50 & 15.57$\pm$0.09 & 14.64$\pm$0.15 & 14.97$\pm$0.25 \\
55090.36 & 15.99$\pm$0.13 & 15.07$\pm$0.12 & 15.19$\pm$0.16 \\
55091.40 & 15.89$\pm$0.15 & 14.95$\pm$0.16 & 14.87$\pm$0.19 \\
55092.36 & 15.99$\pm$0.16 & 15.13$\pm$0.15 & 15.1$\pm$0.2 \\
55095.35 & 15.82$\pm$0.14 & 14.84$\pm$0.09 & 14.88$\pm$0.24 \\
55096.41 & 15.97$\pm$0.13 & 15.06$\pm$0.11 & 14.94$\pm$0.23 \\
55099.42 & 15.85$\pm$0.17 & 14.93$\pm$0.12 & 14.98$\pm$0.14 \\
55102.43 & 15.87$\pm$0.13 & 14.89$\pm$0.14 & 14.99$\pm$0.09 \\
55113.41 & 15.96$\pm$0.21 & 15.04$\pm$0.18 & 14.91$\pm$0.17 \\
55119.41 & 15.69$\pm$0.14 & 14.84$\pm$0.15 & 14.78$\pm$0.18 \\
55122.43 & 15.8$\pm$0.14 & 14.84$\pm$0.11 & 14.8$\pm$0.19 \\
55125.43 & 15.75$\pm$0.12 & 15.18$\pm$0.15 & 15.11$\pm$0.18 \\
55180.13 & 15.55$\pm$0.1 & 14.71$\pm$0.13 & 14.68$\pm$0.14 \\
55182.07 & 15.47$\pm$0.12 & 14.79$\pm$0.11 & 14.61$\pm$0.13 \\
\enddata
%\tablecomments{}
%\tablenotetext{a}{}
\end{deluxetable}

%%%% TABLE 4 - Spectra
\begin{deluxetable*}{lcccccl}
\tablewidth{0pc}\tighten
\tablecaption{Spectroscopic Observations}
\tablehead{
\colhead{Date} &\colhead{Tel./Inst.} &\colhead{Target} &\colhead{day}  
   &\colhead{Exp. (s)} &\colhead{$\lambda$(\AA)} &\colhead{comment}}
\startdata
2009 Sep.\ 10 &Keck/HIRES &U2773-OT  &22 &1800 &6543-7990 &only red lines detected \\
2009 Sep.\ 11 &Keck/HIRES &SN~2009ip &16 &1200 &6543-7990 &only H$\alpha$ detected \\
2009 Sep.\ 21 &Keck/LRIS  &U2773-OT  &34 &300  &3800-5100  &med. resolution \\
2009 Sep.\ 21 &Keck/LRIS  &U2773-OT  &34 &300  &6250-7850  &med. resolution \\
2009 Sep.\ 21 &Keck/LRIS  &U2773-OT  &34 &300  &3600-9200  &low resolution \\
2009 Sep.\ 21 &Keck/LRIS  &SN~2009ip &25 &780  &3800-5100  &med. resolution \\
2009 Sep.\ 21 &Keck/LRIS  &SN~2009ip &25 &640  &6250-7850  &med. resolution \\
2009 Sep.\ 21 &Keck/LRIS  &SN~2009ip &25 &300  &3600-9200  &low resolution \\
\enddata
%\tablecomments{}
%\tablenotetext{a}{}
\end{deluxetable*}

%%%%%%%%%%%%%%%%%%%%%%%%% FIGURE 3 - lightcurves %%%%%%%%%%%%%%%%%%%%%%%
\begin{figure*}
\epsscale{1.05}
%\plotone{../LIGHTCURVE/absmagLC.eps}
\plotone{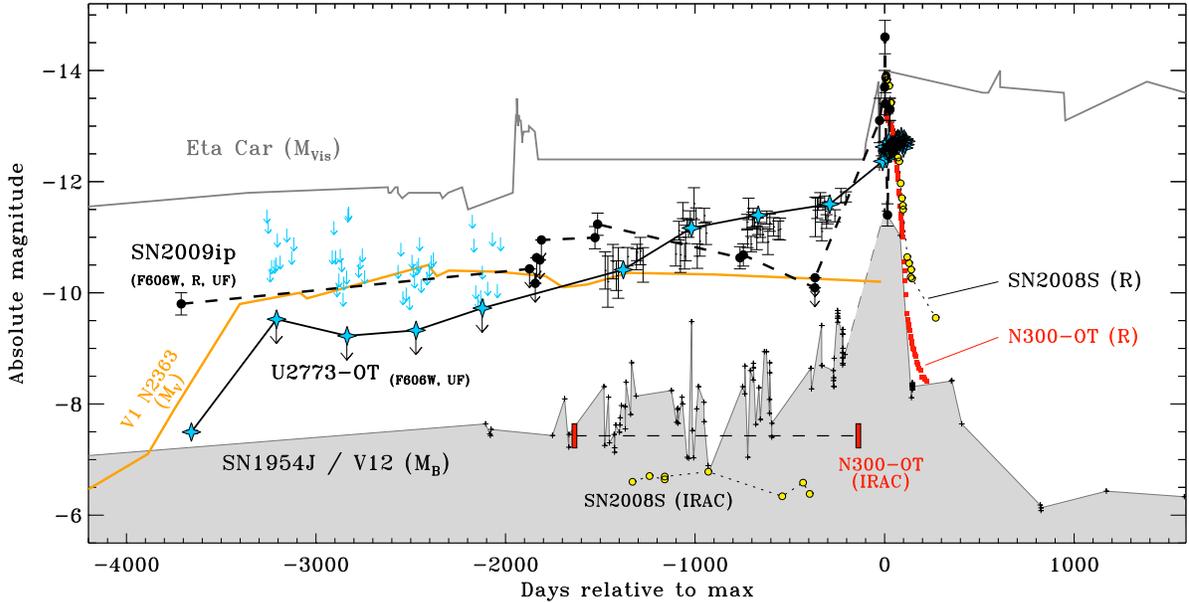}
\caption{Our absolute-magnitude light curves of the LBV-like
  transients SN~2009ip and U2773-OT (Tables 1 and 2) compared to the
  historical LBVs $\eta$~Carinae in its 19th century ``Great
  Eruption'' (compiled by Frew 2004) and SN~1954J (V12 in NGC~2403;
  Tammann \& Sandage 1968).  The orange curve is the LBV eruption of
  V1 in NGC~2363 (Drissen et al.\ 2001; Petit et al.\ 2006), although
  shifted to an arbitrary date.  We also include $R$-band light curves
  of the recent transients N300-OT (from Bond et al.\ 2009) and
  SN~2008S (from Smith et al.\ 2009), along with their pre-outburst
  bolometric magnitudes derived from {\it Spitzer}/IRAC data (Prieto
  et al.\ 2008; Prieto 2008; Bond et al.\ 2009). For U2773-OT, we show
  both upper limits or detections from individual exposures (arrows
  and error bars, respectively), as well as upper limits and
  detections in stacked seasonal data (blue stars).  The earliest
  detections of both objects were from archival {\it HST} F606W data,
  corrected appropriately for Galactic extinction.}
\label{fig:lc}
\end{figure*}
%%%%%%%%%%%%%%%%%%%%%%%%%%%%%%%%%%%%%%%%%%%%%%%%%%%%%%%%%%%%%%%%%%%%%%

\subsection{Pre-Eruption Photometry}

\begin{figure*}
\epsscale{0.95}
%\plotone{../LIGHTCURVE/absmagLC2.eps}
\plotone{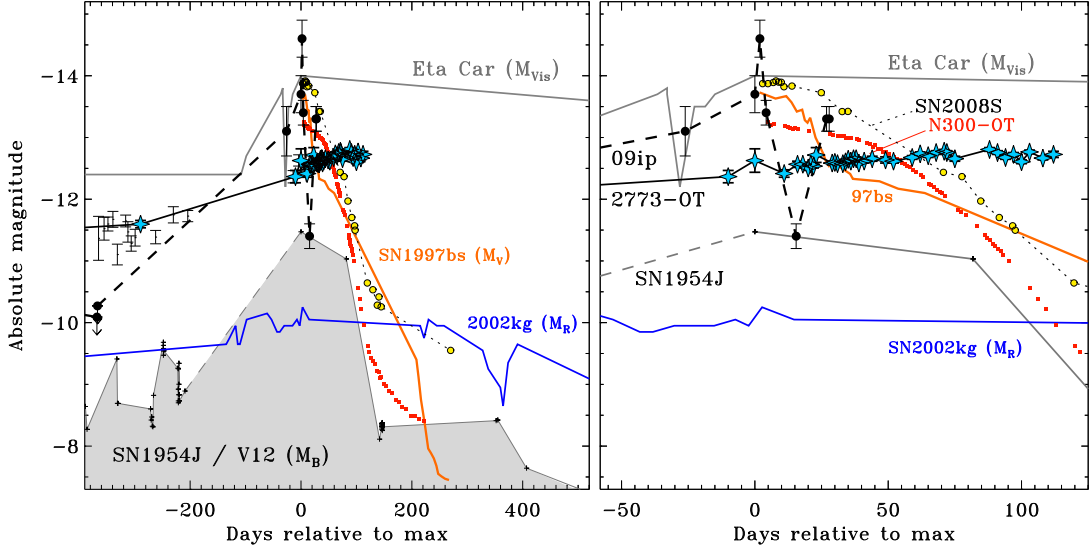}
\caption{Same as Figure~\ref{fig:lc}, but zooming in on the time
  period near maximum light.  Plotting symbols for SN~2009ip, U2773-OT,
  SN~2008S, and N300-OT are the same as in Figure~\ref{fig:lc}.  For
  comparison at this scale, we have also added the $V$-band light
  curve of SN~1997bs (orange; Van Dyk et al.\ 2000) and the
  unfiltered/$R$-band light curve of SN~2002kg (blue; Van Dyk et al.\
  2006).  The right panel shows an expanded view of the left panel.}
\label{fig:lc2}
\end{figure*}

In Figure~\ref{fig:lc} we plot absolute magnitudes derived from the
photometry listed in Tables 1 and 2.  For SN~2009ip in NGC~7259, we
adopt a distance modulus of $m-M=31.55$ mag, and a Galactic reddening
and extinction of $E(B-V) = 0.019$ mag and $A_R = 0.05$ mag,
respectively. Similarly, for U2773-OT we adopt $m-M=28.82$ mag,
$E(B-V) = 0.56$ mag, and $A_R = 1.51$ mag.  The observations are
described below.  The light curves are also shown in
Figure~\ref{fig:lc2}, which focuses on the time around peak
luminosity.

The field of SN~2009ip was imaged multiple times during the operations
of the Palomar-Quest survey, and those observations have been
reprocessed as part of the DeepSky
project\footnote{http://supernova.lbl.gov/{\textasciitilde}nugent/deepsky.html.}
(DS; Nugent 2009). As first noted by Miller et al.\ (2009), a source
at the location of SN~2009ip was observed in 2005, 4~yr prior to
discovery, and it was at comparable luminosity in 2007.  DS typically
has more than one image on any given night when a field was observed,
so we stack all DS images taken on the same night to improve the
limiting magnitude for each epoch. Deep Sky images were obtained
through a nonstandard red filter that has a blue cutoff at $\lambda
\approx$ 6100~\AA, and are otherwise similar to unfiltered photometry,
which we take to be comparable to $R$-band (see below).  DS photometry
of SN~2009ip was calibrated relative to the USNO-B1.0 red magnitudes,
with typical uncertainties of 0.1--0.2 mag when several USNO stars are
used.

The full historical DS light curve of SN~2009ip is shown in
Figure~\ref{fig:lc}, while the photometry is reported in Table~1.
Figures~\ref{fig:lc} and \ref{fig:lc2} also include data around the
time of discovery (Maza et al.\ 2009) and our own $R$-band photometry
obtained with the 0.76-m Katzman Automatic Imaging Telescope (KAIT;
Filippenko et al. 2001; Filippenko 2003) at Lick Observatory (see
below).  The KAIT $R$-band photometry is calibrated relative to the
USNO-B1.0 red magnitudes. It is unclear how the unfiltered discovery
magnitudes reported by Maza et al.\ (2009) are calibrated, so we give
large uncertainties for these particular measurements. Moreover, as
discussed by Li et al.\ (2003), unfiltered data are often clearly
matched to the broad $R$ band.  A source at the position of SN~2009ip
is seen in the {\it HST}/WFPC2 image taken $\sim$10 yr before
discovery, with $m_{F606W} = 21.8 \pm 0.2$ mag.  At the distance of
NGC~7259 this implies $M_V \approx -9.8$ mag.  The early {\it HST}
detection, if this is the quiescent progenitor star, thus requires a
high luminosity of at least log$(L/{\rm L}_{\odot}) \approx 5.9$
(higher for a nonzero bolometric correction), and implies a high
initial mass of 50--80~M$_{\odot}$ (e.g., Lejeune \& Schaerer 2001).

UGC~2773, the dwarf irregular host galaxy of U2773-OT, is monitored
regularly with KAIT. We analyzed prediscovery unfiltered
(approximately $R$ band) images and detected a source at the position
of U2773-OT during the $\sim$5~yr before discovery, as well as upper
limits before that.  There were multiple observations each year, so we
produced stacked seasonal averages to improve the sensitivity.  A
stacked image from the year 2000, when U2773-OT was not detected, is
used as a template image in an image-subtraction technique to cleanly
remove the galaxy contamination at the position of U2773-OT in later
images.  The resulting upper limits and detections are listed in
Table~2, while these averaged data as well as individual epochs are
shown in Figure~\ref{fig:lc}.

As noted above, U2773-OT was also detected in archival {\it HST}/WFPC2
images about 10 yr before discovery, and these {\it HST} magnitudes
are listed in Table~2 as well.  F606W is not a standard $V$-band
filter, so we used SYNPHOT to convert it in order to interpret the
F606W$-$F814W color.  Using SYNPHOT and adopting the Galactic
reddening of $E(B-V) = 0.564$ mag along the line of sight to U2773,
the object has $M_V \approx -7.8$ mag, and an intrinsic $V-I$ color of
$\sim$0.09 mag, or less if there was additional circumstellar
reddening as we strongly suspect (see below).  This corresponds to an
early A-type supergiant or hotter, with log$(L/{\rm L}_{\odot}) \ge
5.1$, and an initial mass of at least 20~M$_{\odot}$.  This is much
like the LBV star HD~168625 (Smith 2007), and it has the same spectral
type but is less luminous than the yellow hypergiant IRC+10420 which,
interestingly, has a spectrum similar to that of the U2773-OT
outburst, as we discuss below.  We find it quite likely, however, that
the progenitor star had some additional circumstellar dust based on
its IR excess (\S 3.5), in which case it was even hotter, more
luminous, and more massive than our estimates from {\it HST} data
alone.

%%%%%%%%%%%%%%%%%%%%%%%%% FIGURE 4 -
%%%%%%%%%%%%%%%%%%%%%%%%% lightcurves %%%%%%%%%%%%%%%%%%%%%%%
%%%%%%%%%%%%%%%%%%%%%%%%%%%%%%%%%%%%%%%%%%%%%%%%%%%%%%%%%%%%%%%%%%%%%%

\subsection{Infrared Photometry During Eruption}

\begin{figure}
\epsscale{1.0}
%\plotone{NIR-u2773.eps}
\plotone{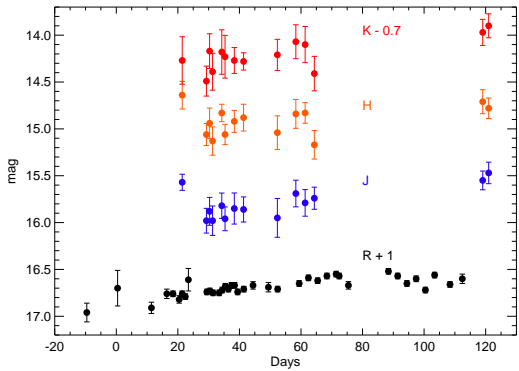}
\caption{The near-IR apparent $JHK_s$ photometry of U2773-OT in days
  after discovery (2009 Aug.\ 18.08) obtained with PAIRITEL, compared
  to the KAIT unfiltered ($\sim$$R$-band) optical photometry.}
\label{fig:ir}
\end{figure}

We observed U2773-OT simultaneously in the $JHK_s$ bands with the
1.3-m Peters Automated Infrared Imaging Telescope (PAIRITEL; Bloom et
al.\ 2006), beginning on 2009 Sep.\ 8 (day 21 after discovery) and at
several subsequent epochs as listed in Table 3. The PAIRITEL
observations were scheduled and obtained automatically by the robotic
telescope, and the images were processed and reduced as part of an
automatic pipeline. Archival Two Micron All Sky Survey (2MASS;
Skrutskie et al. 2006) images of the field taken on 1998 Nov.\ 11 were
used as reference data for image differencing. We performed
PAIRITEL$-$2MASS image subtraction using
HOTPANTS\footnote{http://www.astro.washington.edu/users/becker/hotpants.html.},
and measured the flux of the transient in the difference images via
aperture photometry.  The photometry was calibrated relative to the
2MASS stars in the field.  To estimate the uncertainties we inserted
fake stars at the measured magnitude of U2773-OT on top of locations
in the host galaxy that had a surface brightness similar to that of
the transient's location. We subtracted the 2MASS image from the
images with fake stars, and measured the scatter in the fake-star flux
to determine the uncertainty in the flux of the transient.  The final
$JHK_s$ photometry is listed in Table 3 and is plotted in
Figure~\ref{fig:ir}.  During the time of our observations, the near-IR
flux from U2773-OT has shown a slight increase over $\sim$100 days,
commensurate with the slow brightening in optical photometry, and with
relatively constant color.

Using the stacked $K'$ image of SN~2009ip (\S 2.1), we attempted to
measure its $K'$ magnitude. In a deep PAIRITEL image, in which we do
not detect SN~2009ip, we do detect a star located at $\alpha_{\rm
  J2000}$ = 22$^{\rm h}$23$^{\rm m}$08$\fs$503, $\delta_{\rm J2000}$
=$-$28$\arcdeg$56$\arcmin$47$\farcs$75, which is common to both the
PAIRITEL and AO images. Calibrating relative to 2MASS we measure $K_s
=$ 15.93 $\pm$ 0.12 mag for this star. Assuming $K_s \approx K'$ (the
$K'$ and $K_s$ filters are approximately the same), we therefore
measure $K' = 19.68 \pm 0.12$ mag for SN~2009ip on 2009 Sep.\ 9.

%%%%%%%%%%%%%%%%%%%%%%%%% FIGURE 5 -
%%%%%%%%%%%%%%%%%%%%%%%%% optical / IR phot of U2773 %%%%%%%%%%%%%%%%%%%%%%%
%%%%%%%%%%%%%%%%%%%%%%%%%%%%%%%%%%%%%%%%%%%%%%%%%%%%%%%%%%%%%%%%%%%%%%

%%%%%%%%%%%%%%%% spectra
\subsection{Spectroscopy}

A summary of our spectroscopic observations is listed in Table 4.  We
obtained high-resolution ($R = \lambda/\Delta\lambda \approx 60,000$)
optical echelle spectra of U2773-OT on 2009 Sep.\ 10 using HIRES (Vogt
et al. 2004) on the 10-m Keck~I telescope. The HIRES spectra were
reduced using standard procedures.  These observations correspond to
day 22 after discovery for U2773-OT.  We used the B5 decker
(0$\farcs$86 slit width) and a total exposure time of 1800~s.  The
instrument covers the wavelength range 3642--7990 \AA, but with gaps
in the wavelength coverage because the spectrum is dispersed onto
three different detectors; because of the low signal-to-noise ratio
(S/N), we only used data on the red chip over the interval
6543--7990~\AA. This was a single exposure, so individual cosmic rays
in the extracted spectrum were masked out; these features were always
a few pixels wide and did not significantly impact the line profiles.

The spectrum has very low S/N in the blue range, but several emission
lines are detected at red wavelengths.  The spectrum reveals narrow
emission lines of H$\alpha$, [N~{\sc ii}] $\lambda\lambda$6548, 6583,
and [S~{\sc ii}] $\lambda\lambda$6716, 6731 having full width at
half-maximum intensity (FWHM) $\approx$ 49 km s$^{-1}$ (much wider
than the instrument resolution of $\sim$5 km s$^{-1}$); these are due
to an underlying H~{\sc ii} region that was not subtracted, or perhaps
an extended circumstellar nebula.  In addition, H$\alpha$ has a
broader base attributable to the transient.  The HIRES spectrum of
U2773-OT in the wavelength range around H$\alpha$ is shown in
Figure~\ref{fig:hires1}.  The underlying broad feature has a clear
P~Cygni profile, which can be fit with a combination of an emission
line having Gaussian FWHM $\approx$ 360 km s$^{-1}$ and another
Gaussian in absorption.  The P~Cygni absorption trough is at $-$350 km
s$^{-1}$, in agreement with velocities quoted by Berger \& Foley
(2009) in an earlier spectrum.

We obtained a HIRES spectrum of SN~2009ip on the following night, 2009
Sep.\ 11 (day 16).  Despite good conditions, the 20-min exposure was
not deep enough to detect the continuum, and subsequent analysis of
unfiltered photometry of the guider image revealed that SN~2009ip had
faded more than we anticipated, to 20.2 mag.  Our HIRES observation
was conducted at or near the minimum of a sudden fading episode,
discussed at length in \S 3.2.  At this epoch, however, we did detect
the H$\alpha$ emission line, which has a total line flux of $0.7
\times 10^{-14}$ erg s$^{-1}$ cm$^{-2}$ (this is one third the value
measured in lower-resolution spectra obtained a few days later; see
below).  The HIRES H$\alpha$ line profile in SN~2009ip is shown in
Figure~\ref{fig:hires2}; the spectrum is very noisy, even though the
pixels have been binned by a factor of 4.  Given the quality of the
data, the line can be approximated adequately by a Lorentzian profile
with FWHM $\approx$ 550 km s$^{-1}$; this is the same profile that is
well fit to the lower-resolution spectrum (see below).  We do not
clearly detect any narrow component from an underlying H~{\sc ii}
region around SN~2009ip, although this is not necessarily surprising
as many nearby LBVs are not in bright H~{\sc ii} regions.

Later, on 2009 Sep.\ 21, we observed SN~2009ip again, this time with
the Low Resolution Imaging Spectrometer (LRIS; Oke et al.\ 1995) on
Keck~I.  To our surprise, unfiltered photometry in the guider image
revealed that SN~2009ip had rebrightened to 18.3 mag (listed in Table
1), returning almost to its peak brightness.  The conditions were
photometric, with 0$\farcs$85 seeing, so we obtained spectra of both
SN~2009ip and U2773-OT.  We observed with the same LRIS setup for both
SN~2009ip and U2773-OT, consisting of medium-resolution (0$\farcs$7
slit width; 1200 lines mm$^{-1}$ grating; 3~\AA\ pixel$^{-1}$) and
low-resolution (1$\farcs$0 slit width; 400 lines mm$^{-1}$ grating;
$\sim$6~\AA\ pixel$^{-1}$) red-side settings. On the blue side, a 400
lines mm$^{-1}$ grism was used in both settings, providing spectra
with $\sim$3~\AA\ pixel$^{-1}$. The medium-resolution setting yielded
blue and red spectra in the wavelength ranges 3120--5594~\AA\ and
6250--7880~\AA, respectively, while the low-resolution setting covered
the full range 3246--10,260~\AA\ (the bluest wavelengths were clipped
due to noise).  For SN~2009ip, we used exposure times of 780~s (blue
side, medium resolution), $2 \times 320$~s (red side, medium
resolution), and 300~s (both sides, low resolution).  For U2773-OT,
all exposures were 300~s.

The reduction of the U2773-OT spectrum was complicated because the
long-slit spectra revealed extended emission along the slit due to a
background H~{\sc ii} region or extended circumstellar nebula within a
few tenths of an arcsecond from the transient.  This background H~{\sc
  ii} region emission was sampled on either side of the source and
carefully subtracted from the spectrum, although some subtraction
residuals remained at low levels (see, for example, the [S~{\sc ii}]
lines in the red-side spectrum).

Figures~\ref{fig:lrisHB} and \ref{fig:lrisHR} show the final
medium-resolution LRIS spectra of both transients in the blue and red,
respectively, while Figure~\ref{fig:lrisHa} zooms in on the H$\alpha$
profiles of each.  Figure~\ref{fig:ca2} displays the wavelength range
around the red [Ca~{\sc ii}] lines seen in the LRIS spectra for both
targets, as well as the HIRES data for U2773-OT.  Figure~\ref{fig:nad}
shows the wavelength range around Na~{\sc i} D and He~{\sc i}
$\lambda$5876 in the low-resolution LRIS spectra, which is in a
spectral region excluded by the medium-resolution LRIS data.  Finally,
the full low-resolution LRIS spectra of both SN~2009ip and U2773-OT
are shown in Figure~\ref{fig:spec}.  In Figure~\ref{fig:spec} we
compare these low-resolution LRIS spectra of both transients to
spectra of the yellow hypergiant IRC+10420 (from Smith et al.\ 2009),
as well as SN~2008S (Smith et al.\ 2009) and the LBV SN~1997bs (Van
Dyk et al.\ 2002) during eruption.  The low-resolution LRIS spectrum
of U2773-OT was qualitatively similar, although superior to, a
spectrum obtained at Lick Observatory three nights earlier, which is
not shown here.

\section{RESULTS}

\subsection{Light-Curve Comparison}

Absolute-magnitude light curves for SN~2009ip and U2773-OT are
displayed in Figure~\ref{fig:lc}, where they are compared with those
of several other objects.  The gray line shows the historical visual
light curve of $\eta$~Car during its 19th century ``Great Eruption,''
compiled from historical sources by Frew (2004).  It showed gradual
brightening 10~yr before peak, a pre-outburst event in 1837, and then
finally the $M_{\rm Visual} \approx -14$ mag peak of its eruption in
1843, followed by a slow and irregular decline over the next decade.
While $\eta$~Car is the most famous and best studied LBV, it is
certainly not representative of all LBVs.  The prolonged eruption, in
particular, is highly unusual.

We also show the $B$-band light curve of the prototypical LBV eruption
SN~1954J (V12 in NGC~2403; shaded gray in Figure~\ref{fig:lc}) from
Tammann \& Sandage (1968), corrected for Galactic extinction.  Before
the LBV eruption began, the quiescent star was considerably less
luminous than $\eta$~Car, with an absolute $B$ magnitude of only about
$-7$ to $-7.5$ mag.  Yet, Tammann \& Sandage (1968) noted that it was
a key example of the class of bright blue irregular variables like
those in M31 and M33 (Hubble \& Sandage 1953), later termed ``LBVs,''
and the star is now known to have survived the event (Smith et al.\
2001; Van Dyk et al.\ 2005).  The progenitor, V12, is a member of the
less luminous class of LBVs with initial masses of 20--40 M$_{\odot}$
(Smith et al.\ 2004).  V12 showed rapid irregular variability in the
$\sim$5 yr leading up to its outburst, with oscillations of 1--2 mag.
(This ``flickering'' is discussed more in \S 3.2 below.) Such rapid
variability is unusual, but V12 also has unusually well-sampled
pre-eruption photometry.  The cause of these wild oscillations is
unknown, but they probably signify a growing instability in the star
and herald the approaching runaway of the SN~1954J event
itself.\footnote{While the peak of the SN~1954J eruption was fainter
  than others in Figure~\ref{fig:lc}, it is worth noting that it was
  not observed for $\sim$7 months before peak, and Tammann \& Sandage
  (1968) suspected that a brighter peak magnitude may have occurred
  during that hiatus from observing.  We indicate this time interval
  with a dashed line in Figs.~\ref{fig:lc} and \ref{fig:lc2}.}

Another case to consider is the more recent and well-studied LBV
outburst V1 in the nearby dwarf irregular galaxy NGC~2363.  The
absolute $V$ light curve from Drissen et al.\ (1997, 2001; see also
Petit et al.\ 2006) is shown with the orange curve in
Figure~\ref{fig:lc}.  (Its dwarf host with metallicity similar to that
of the Small Magellanic Cloud is particularly relevant for U2773-OT.)
V1 is noteworthy because, again, its LBV eruption came from a star
whose initial luminosity was low compared to that of $\eta$ Car --- in
fact, its quiescent pre-outburst luminosity was equivalent to the IR
luminosity of SN~2008S, also shown in Figure~\ref{fig:lc} (more
details are given below).  During its decade-long outburst in the
1990s, its $M_V$ brightened by $\sim$3.5 mag and remained so for
several years.  Unlike the historical examples of V12 and $\eta$ Car,
V1 has been subject to detailed non-LTE (local thermodynamic
equilibrium) modeling of its outburst spectrum, revealing log$(L/{\rm
  L}_{\odot}) \approx 10^{6.4}$ and $R/{\rm R}_{\odot} \approx$
300--400 during its outburst, and a strong stellar wind with
$v_{\infty} \approx 300$ km s$^{-1}$ (Drissen et al.\ 2001).  The
radius of the photosphere, however, only increased at a rate of
$\sim$4 km s$^{-1}$ (i.e., much slower than the steady wind
expansion), so this was not an explosion.  V1 lives within the
well-known ``mini-starburst'' giant H~{\sc ii} region NGC~2366
(Drissen et al.\ 1997) in the galaxy NGC~2363, surrounded by many
young, massive stars.  This is a well-established case of a
super-Eddington wind outburst from a massive star, despite the
apparently low luminosity of its quiescent-phase progenitor. The LBV
outburst of V1 is similar in duration and absolute magnitude to the
precursor outbursts of the two new transients presented in this paper,
although the V1 LBV outburst has not (as yet) culminated in a
comparably bright giant eruption phase with $M_V \la -12$ mag;
instead, it appears to be an example of a fainter and prolonged LBV
eruption.

Both SN~2009ip and U2773-OT stand in between $\eta$~Car and SN~1954J,
with all four objects showing precursor variability in the decade
leading up to the peak of their giant eruptions.  Again, the cause for
this is not known, but the phenomenon is a well-established property
of some LBVs (e.g., Humphreys et al.\ 1999).  This precursor
variability and the range of luminosity of the progenitors makes a
strong case that SN~2009ip and U2773-OT are, in fact, both {\it bona
  fide} giant LBV eruptions from massive stars that were in a
prolonged outburst phase before their discovery.

Figure~\ref{fig:lc} also includes some available information for
SN~2008S and N300-OT, and their dust-obscured progenitors.  The light
curves of their eruptions show peak absolute magnitudes of roughly
$-$14 mag, with a relatively fast decline over 100--200 days
resembling that of SN~1954J.  Their similarity to LBV eruptions was
already noted in previous papers (Smith et al.\ 2009; Berger et al.\
2009a; Bond et al.\ 2009; Prieto et al.\ 2008; Thompson et al.\ 2009),
but they were surprising because of their relatively low-luminosity
progenitors compared to LBVs.  However, Figure~\ref{fig:lc} shows that
{\it their progenitor luminosities are not that low after all}.  The
plotted quantities in Figure~\ref{fig:lc} are the bolometric
luminosities derived from fits to the mid-IR spectral energy
distributions (SEDs) measured in {\it Spitzer} data (Prieto et al.\
2008; Prieto 2008; Bond et al.\ 2009).  While their luminosities may
overlap with the most extreme AGB stars at the very tip of their
evolution (Thompson et al.\ 2009), the luminosity of N300-OT is the
same as the pre-outburst luminosity of V12 in NGC~2403, known to be an
LBV, and is very close to the pre-outburst luminosity of U2773-OT.
The IR luminosity of the SN~2008S progenitor is only a factor of
$\sim$2 less, and is comparable to the quiescent luminosity of the LBV
V1 in NGC~2363.  This makes it plausible that both of these transients
were moderately massive stars, comparable to or somewhat less massive
than V12.  This LBV connection is reinforced by a spectral comparison,
discussed later.

An obvious caveat is that we are comparing integrated IR luminosities
of dust-enshrouded progenitors (for SN~2008S and N300-OT) to estimates
of the luminosity based on visual-wavelength photometry. The
comparison in Figure~\ref{fig:lc} assumes bolometric corrections of
zero for the optically identified sources, and does not include
possible IR excesses or correction for local extinction, so these
luminosities are actually lower limits.  However, the integrated mid-IR
luminosities of SN~2008S and N300-OT are also lower limits, since they
only represent the luminosity absorbed and reradiated by warm dust,
whereas radiation may escape in other directions that we cannot see
without heating dust if the dust shells are nonspherical (in the case
of an edge-on dust torus, for example).  In any case, further
corrections beyond the values shown in Figure~\ref{fig:lc} become very
uncertain, but could only raise the luminosities shown here.

\subsection{Peak Luminosity, Decline Rates, and ``Flickering''}

The photometric behavior around the time of peak luminosity is unusual
for both transients.  Figure~\ref{fig:lc2} shows the same light curves
as in Figure~\ref{fig:lc}, but on an expanded scale to illustrate the
details of the giant eruptions themselves.

In the 5 yr before discovery, U2773-OT was apparently in an unstable
pre-outburst state. It continually rose in brightness, culminating in
its peak absolute $R$ magnitude of about $-$12.8.  The total increase
was ${\Delta}m \approx 5$ mag.  It has remained roughly at that
luminosity for a few months afterward, showing minor oscillations with
amplitudes of no more than 0.2 mag on timescales of several days.

SN~2009ip was different and rather astonishing.  After its precursor
eruption $\sim$5 yr before discovery, it settled to a fainter state,
with only upper limits of about $-$10 mag at $\sim$1 yr before
discovery.  This may mark a temporary return to its quiescent state
seen by {\it HST}.  It then brightened to $M_R = -13$ mag, and
continued to rise to a peak magnitude even brighter than that of
$\eta$~Car, at $M_R = -14.5$ mag.  The total increase in brightness
from its pre-outburst state was $\sim$4.7 mag.  Unexpectedly, however,
SN~2009ip suddenly faded by 3.2~mag in $\sim$16~days, only to recover
again soon after (Fig.~\ref{fig:lc2}).  We announced our discovery of
this startling dip and recovery shortly before submitting this paper
(Li et al.\ 2009).

%%%%%%%%%%%%%%%%%%%%%%%%% FIGURE 6 - hires spec %%%%%%%%%%%%%%%%%%%%%%%%%%%
\begin{figure}
\epsscale{0.99}
%\plotone{../SPECTRA/halpha.eps}
\plotone{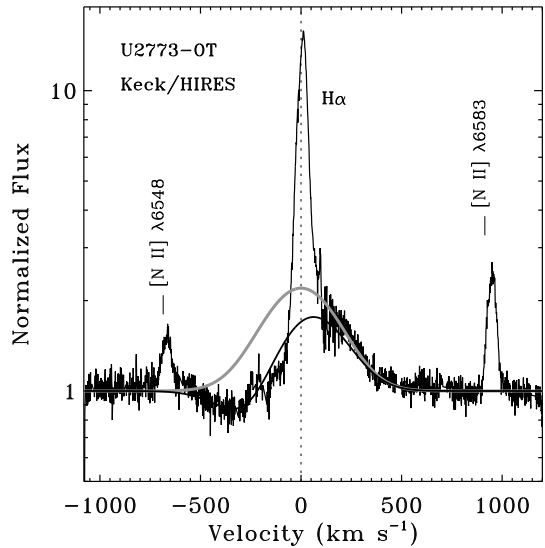}
\caption{HIRES spectrum of U2773-OT on day 22, showing H$\alpha$ and
  the narrow [N~{\sc ii}] lines.  The thick gray curve is a Gaussian
  with FWHM = 360 km s$^{-1}$, and the thin black curve is the same
  Gaussian with a blueshifted P~Cygni absorption component subtracted.
  The absorption minimum is at $-$350 km s$^{-1}$.  The narrow
  components of all lines have Gaussian FWHM $\approx$ 49 km s$^{-1}$,
  and arise in an underlying H~{\sc ii} region or extended
  circumstellar nebula that was not subtracted from the HIRES data.}
\label{fig:hires1}
\end{figure}
%%%%%%%%%%%%%%%%%%%%%%%%%%%%%%%%%%%%%%%%%%%%%%%%%%%%%%%%%%%%%%%%%%%%%%%
%%%%%%%%%%%%%%%%%%%%%%%%% FIGURE 7 - hires spec %%%%%%%%%%%%%%%%%%%%%%%%%%%
\begin{figure}
\epsscale{0.99}
%\plotone{../SPECTRA/halpha09ip.eps}
\plotone{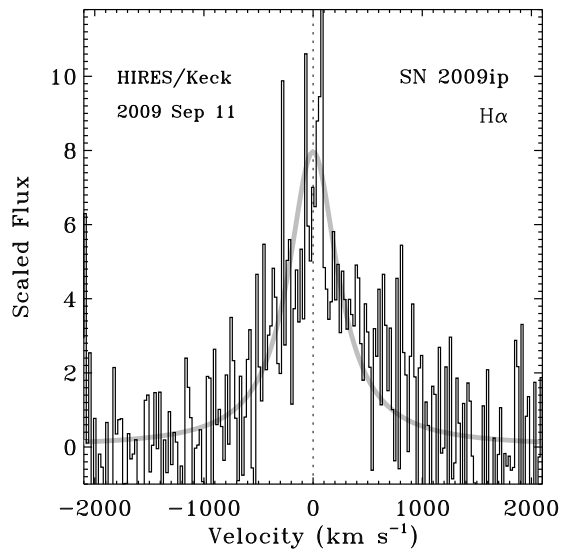}
\caption{HIRES spectrum of SN~2009ip on day 16, showing H$\alpha$.
  The spectrum has been binned by a factor of 4 to reduce noise, and
  the continuum was not clearly detected.  The thick gray curve is a
  Lorentzian profile with FWHM = 550 km s$^{-1}$.}
\label{fig:hires2}
\end{figure}
%%%%%%%%%%%%%%%%%%%%%%%%%%%%%%%%%%%%%%%%%%%%%%%%%%%%%%%%%%%%%%%%%%%%%%%

In connection with the transients SN~2008S and N300-OT, one may wonder
if their fast decline rates are consistent with LBVs.  In fact,
several well-studied LBVs have shown extremely fast decline rates.  At
$\sim$100~days after peak, V12/SN~1954J exhibited a very rapid decline
rate of 0.05 mag day$^{-1}$, comparable to those of SN~2008S and N300-OT
(Fig.~\ref{fig:lc2}).  SN~1997bs also showed an extremely rapid 1.5
mag decline from its peak magnitude in the first 20 days, although
that decline rate varied later.  Even $\eta$~Car had a rapid rise and
decline associated with its events in 1837 and 1843
(Fig.~\ref{fig:lc}).

%%%%%%%%%%%%%%%%%%%%%%%%%% FIGURE 8 - medres spec %%%%%%%%%%%%%%%%%%%%%%%%%%%
\begin{figure*}
\epsscale{0.9}
%\plotone{../SPECTRA/lrisHB.eps}
\plotone{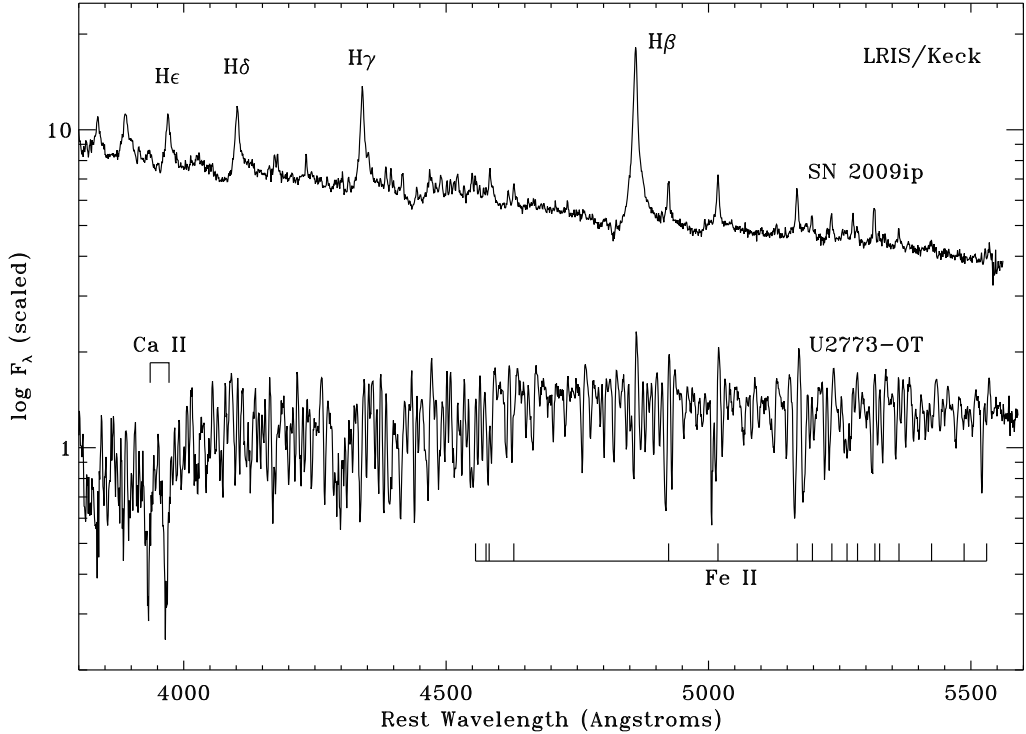}
\caption{Medium-resolution LRIS spectra of SN~2009ip (day 25) and
  U2773-OT (day 34) in the blue, dereddened by $E(B-V)$ values of 0.019
  mag and 0.564 mag for SN~2009ip and U2773-OT, respectively. The forest
  of absorption lines in U2773-OT is real.}
\label{fig:lrisHB}
\end{figure*}
%%%%%%%%%%%%%%%%%%%%%%%%%%%%%%%%%%%%%%%%%%%%%%%%%%%%%%%%%%%%%%%%%%%%%%%

In this context, the astonishingly fast decline of SN~2009ip from its
maximum luminosity is quite important.  Its prediscovery luminosity
and variability establish that it is a true LBV, yet it fades faster
than any of the historical LBVs, SN impostors, or the controversial
transients SN~2008S and N300-OT.  The sharp fading and rebrightening
of SN~2009ip is even more extreme than the fluctuations experienced by
$\eta$ Car in 1837 and 1843, providing some assurances that the rapid
19th century fluctuations observed by J. F. W.\ Herschel may have, in
fact, been real, leading him (see Herschel 1847) to describe
$\eta$~Car as ``a star fitfully variable to an astonishing
extent... apparently with no settled period and no regularity of
progression.''

To continue quoting Herschel: ``What origin can we ascribe to these
sudden flashes and relapses?''  This old mystery persists, and is made
even more extreme by the case of SN~2009ip.  A fading of over 3 mag in
16 days is too fast for most physical mechanisms one can imagine, and
is faster than most SNe.  It cannot be a huge increase in extinction
caused by a simple puff of dust formation, since the time for ejected
material to reach the dust sublimation radius of 170--230 AU (for the
observed luminosity of $\sim2 \times 10^7$ L$_{\odot}$) expanding at
$\sim$500 km s$^{-1}$ is much longer --- roughly 600--800~days
depending on grain-condensation temperatures.  We can rule out
substantial dust anyway, based on the lack of IR excess during the
fading, as discussed below in \S 3.5.  Furthermore, after the dip, the
luminosity of SN~2009ip recovered faster than can be explained by the
subsequent thinning of that hypothetical dust.  One can imagine that
hydrogen in the high-density wind could recombine quickly, but this
requires that the source of ultraviolet (UV) photons was suddenly
quenched, and it seems inconsistent with our detection of H$\alpha$
during the dip.  Unless it was much hotter than typical LBV eruptions,
the photospheric radius of SN~2009ip must have been comparable to the
orbit of Saturn, but fluctuating as fast as (or faster than) the
wind's expansion speed, so the sudden ejection of an optically thick
shell is perhaps the most likely culprit.  Davidson \& Humphreys
(1997) noted that in the case of $\eta$~Car, this rapid fluctuation
challenges even the dynamical timescale of the star itself.  A fading
by more than a factor of 10 over such a short time in SN~2009ip is
truly spectacular.

Other LBVs show qualitatively similar fading and rebrightening
episodes, although less extreme, which we refer to as ``flickering.''
V12 in NGC~2403, for example, oscillated wildly in the decade before
its eruption (Fig.~\ref{fig:lc}), as noted above, with several changes
of more than 1 mag on equally short timescales (Tammann \& Sandage
1968).  In the same galaxy, V37/SN~2002kg had a rapid fading and
rebrightening episode about 350--380 days after peak, as plotted in
Figure~\ref{fig:lc2} (see Van Dyk et al.\ 2006).  This sort of rapid
variability argues that short cadences are valuable when obtaining
photometry and spectroscopy of these objects, lest one miss a
significant mass-loss event.  In that case, our method of stacking
seasonal data for the prediscovery variability of U2773-OT may have
been an oversimplification, which is why we also plot individual
measurements in Figures~\ref{fig:lc} and \ref{fig:lc2}.

%%%%%%%%%%%%%%%%%%%%%%%%%% FIGURE 9 - medres spec %%%%%%%%%%%%%%%%%%%%%%%%%%%
\begin{figure*}
\epsscale{0.9}
%\plotone{../SPECTRA/lrisHR.eps}
\plotone{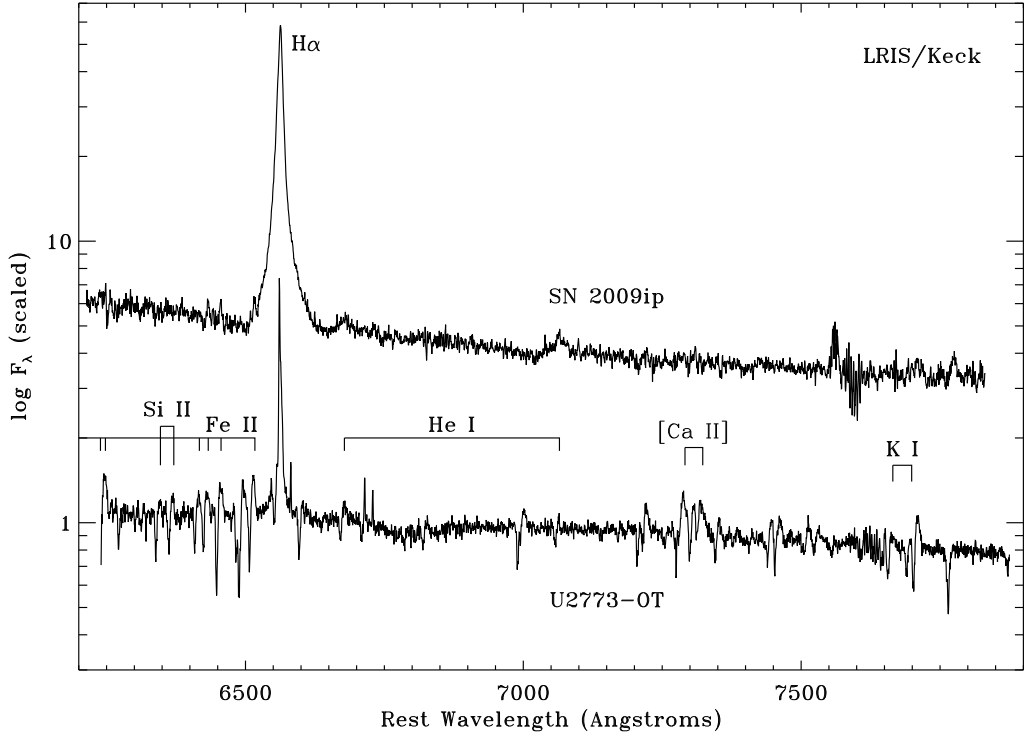}
\caption{Medium-resolution LRIS spectra of SN~2009ip (day 25) and
  U2773-OT (day 34) in the red, dereddened by $E(B-V)$ values of 0.019
  mag and 0.564 mag for SN~2009ip and U2773-OT, respectively.}
\label{fig:lrisHR}
\end{figure*}
%%%%%%%%%%%%%%%%%%%%%%%%%%%%%%%%%%%%%%%%%%%%%%%%%%%%%%%%%%%%%%%%%%%%%%%

\subsection{Spectral Morphology}

Figures~\ref{fig:lrisHB} and \ref{fig:lrisHR} show the
medium-resolution Keck/LRIS spectra at blue and red wavelengths,
respectively, of SN~2009ip (day 25) and U2773-OT (day 34).  These are
useful for discussing the general appearance of the spectra, and the
differences between the two transients.  We describe the spectral
morphology in detail below; in brief, SN~2009ip resembles typical
spectra of LBVs in their hotter state, whereas U2773-OT is exemplary
of the complex spectra of LBVs in their cooler state.

The day 25 spectrum of SN~2009ip in Figures~\ref{fig:lrisHB} and
\ref{fig:lrisHR} is dominated by strong Balmer lines with Lorentzian
FWHM widths of $\sim$550 km s$^{-1}$ and a smooth continuum with an
apparent blackbody temperature of $\sim$10$^4$~K.  It also shows broad
emission profiles of He~{\sc i} $\lambda\lambda$5876, 6678, and 7065
with similar widths (note that He~{\sc i} $\lambda$5876 was in the gap
between the blue and red LRIS medium-resolution spectra in
Figs.~\ref{fig:lrisHB} and \ref{fig:lrisHR}, but it can be seen in the
low-resolution spectrum in Fig.~\ref{fig:nad}).
% The presence of He~{\sc i} lines is interesting because they usually
% signify temperatures of roughly 20,000~K, which is substantially
% hotter than the apparent continuum temperature.
SN~2009ip also exhibits several narrower emission lines which are
mostly Fe~{\sc ii}, probably produced in the outer wind.  Overall, the
spectrum is typical of classical LBVs in their hotter states (e.g.,
Hillier et al.\ 2001; Szeifert et al.\ 1996; Stahl et al.\ 1993; Stahl
1986).  In the dereddened low-resolution day 25 spectrum of SN~2009ip
(Fig.~\ref{fig:spec}), we measure a Balmer decrement of H$\alpha$ :
H$\beta$ : H$\gamma$ = 2.74 : 1.0 : 0.48.  This is very close to Case
B recombination values, and is similar to the decrement H$\alpha$ :
H$\beta$ : H$\gamma$ = 2.6 : 1.0 : 0.5 observed in SN~1997bs (Van Dyk
et al.\ 2000) in a spectrum taken shortly after discovery.  Note,
however, that the spectral evolution of well-studied SNe~IIn such as
SN~1994W and SN~2006gy (Chugai et al.\ 2004; Smith et al.\ 2010) shows
that the Balmer decrement is highly time dependent, steepening as the
emitting layer expands, thins, and cools.  This may be the case in
SN~2009ip as well.

By contrast, the day 34 spectrum of U2773-OT is more complicated, with
narrower emission lines, numerous narrow blueshifted absorption lines,
and a cooler apparent temperature of $\sim$7000~K (this is corrected
only for Galactic reddening, so the intrinsic temperature may be
somewhat warmer).  The prominent absorption indicates that the
eruption wind of U2773-OT has lower ionization than that of
SN~2009ip. Like SN~2008S and N300-OT, it exhibits strong and narrow
emission from [Ca~{\sc ii}] $\lambda$7291 and $\lambda$7323, as well
as bright narrow emission and P~Cygni absorption in the near-IR
Ca~{\sc ii} triplet (the near-IR Ca~{\sc ii} triplet is shown in
Fig.~\ref{fig:spec}), plus strong absorption in Ca~{\sc ii} H and K
(Fig.~\ref{fig:spec}).  He~{\sc i} lines, if present, are extremely
weak.  Among the unusual low-ionization P~Cygni emission features
present in the spectrum of U2773-OT are the K~{\sc i} $\lambda$7665
and $\lambda$7699 resonance lines, rarely seen in emission except in
cases such as the extreme supergiant VY CMa (e.g., Smith 2004, and
references therein).  (Note that K~{\sc i} $\lambda$7665 probably
suffers heavier telluric absorption than $\lambda$7665.)  Except for
H$\alpha$, the higher-order Balmer lines are not prominent, while the
overall appearance of the spectrum is dominated by a dense forest of
narrow absorption lines in the blue, mostly Fe~{\sc ii} and other
low-ionization metal lines; many of the same lines are seen in
emission in SN~2009ip.  Interestingly, many of the same
emission/absorption features are identified in the spectrum of the
Type IIn SN~2006gy (Smith et al.\ 2010), from which the line
identifications in Figures~\ref{fig:lrisHB} and \ref{fig:lrisHR} have
been taken, although the lines in that object are broader.  Similarly,
many of the same lines are seen in the spectrum of IRC+10420
(Humphreys et al.\ 2002), although in that object the outflow is
slower.

Altogether, the spectrum of U2773-OT is a composite of an
emission-line wind spectrum and a dense absorption spectrum of an F
supergiant, as is characteristic for LBVs in their cool eruptive
states.  In fact, its spectrum is an apparent carbon copy of the
spectra of R~127 and S~Doradus in their cool eruptive states (Wolf
1989; Wolf \& Stahl 1990; Wolf et al.\ 1988; Walborn et al.\ 2008),
for which detailed models suggest a temperature of $\sim$8000~K.
Armed with this spectrum of U2773-OT and no other data, an informed
spectroscopist would conclude that it is most likely an LBV in a cool
S~Dor phase (see, e.g., Humphreys \& Davidson 1994).

Given the stark differences displayed by the spectra of SN~2009ip and
U2773-OT in Figures~\ref{fig:lrisHB} and \ref{fig:lrisHR}, how can
they both be LBVs?  The LBVs are a heterogeneous group, but they are
known for the duality of their hot (usually quiescent) and cool
(outburst) S~Dor states.  These two preferred states are demonstrated
well by each of the two transients in Figures~\ref{fig:lrisHB} and
\ref{fig:lrisHR}.  Interestingly, one can find examples where the same
star observed at different points in its variability cycle had a
spectrum that resembled that of either SN~2009ip or U2773-OT, such as
the case of R~127 (Wolf 1989; Walborn et al.\ 2008).  It may seem
puzzling, then, why SN~2009ip is in a hotter spectroscopic state when
photometry of the object clearly shows it in mid-eruption.  While this
behavior is well established for normal S~Dor outbursts, we do not yet
fully understand the spectroscopic behavior of giant LBV eruptions
because spectra of true giant eruptions of LBVs are rare.  SN~2009ip
clearly indicates that an LBV can be hotter than 8000~K during a giant
eruption, as is apparently the case for SN~1997bs (Van Dyk et al.\
2002) and V1 in NGC~2363 (Drissen et al.\ 2001).

\subsection{Emission-Line Profiles and Outflow Speeds}

\subsubsection{U2773-OT}

Figure~\ref{fig:hires1} shows a portion of the spectrum surrounding
H$\alpha$ from the Keck/HIRES observation of U2773-OT obtained on day
22 after discovery.  The spectrum clearly shows narrow components of
H$\alpha$ and the red [N~{\sc ii}] lines, plus a broader underlying
P~Cygni profile in H$\alpha$.  The total equivalent width of H$\alpha$
(narrow plus broad) is only $-25.4 \pm 0.3$~\AA.

\begin{figure}
\epsscale{0.99}
%\plotone{../SPECTRA/halphaLRIS.eps}
\plotone{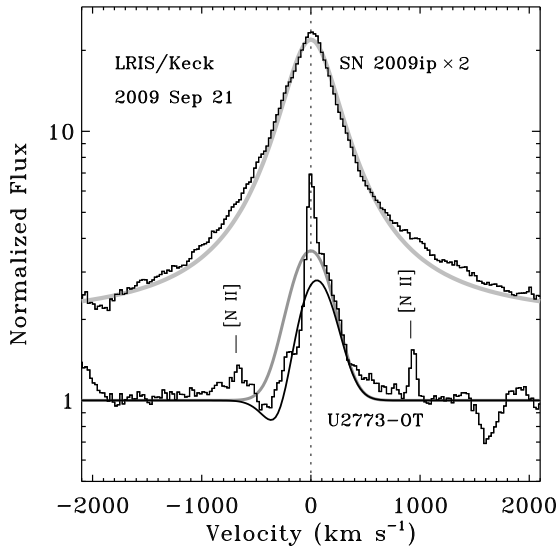}
\caption{Same as Figure~\ref{fig:hires1}, but with the
  medium-resolution LRIS spectra of both U2773-OT and SN~2009ip.  The
  Gaussian curves for U2773-OT are the same velocities as in
  Figure~\ref{fig:hires1}, although with somewhat higher intensities.
  The gray curve matched to the H$\alpha$ profile of SN~2009ip is a
  Lorentzian profile with FWHM = 550 km s$^{-1}$.}
\label{fig:lrisHa}
\end{figure}

The underlying broad H$\alpha$ P~Cygni component can be approximated
by an emission component with Gaussian FWHM $\approx$ 360 km s$^{-1}$
(gray curve in Fig.~\ref{fig:hires1}), plus a Gaussian absorption
component that causes the P~Cygni minimum at $-$350 km s$^{-1}$.  This
assessment based on the day 22 HIRES spectrum is consistent with the
H$\alpha$ profile of U2773-OT observed on day 34 with LRIS
(Fig.~\ref{fig:lrisHa}).  This weak underlying broad component also
matches the expansion speeds observed by Berger \& Foley (2009) about
8 days earlier, although they noted this characteristic speed in the
P~Cygni absorption in the near-IR Ca~{\sc ii} triplet and the full
emission FWHM of H$\alpha$.  We take 350 km s$^{-1}$ to be the
expansion speed of the eruption wind from U2773-OT.  It is a typical
speed for the winds of blue supergiants and LBV stars, but would be
astonishingly slow for any conventional explosion scenario.  In
particular, it is comparable to the 300~km s$^{-1}$ wind observed
spectroscopically in the LBV eruption V1 in NGC~2363 (Drissen et al.\
2001), which we mentioned earlier, as well as 370 km s$^{-1}$ in
V37/SN~2002kg (Van Dyk et al.\ 2006).

%%%%%%%%%%%%%%%%%%%%%%%%% FIGURE 910- LRIS spec %%%%%%%%%%%%%%%%%%%%%%%%%%%
%%%%%%%%%%%%%%%%%%%%%%%%%%%%%%%%%%%%%%%%%%%%%%%%%%%%%%%%%%%%%%%%%%%%%%%

%%%%%%%%%%%%%%%%%%%%%%%%% FIGURE 11 - [Ca II] %%%%%%%%%%%%%%%%%%%%%%%%%%%
\begin{figure}
\epsscale{0.99}
%\plotone{../SPECTRA/ca2.eps}
\plotone{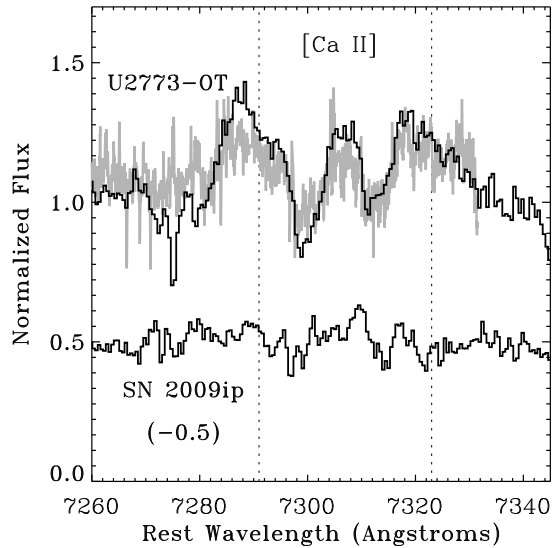}
\caption{LRIS spectra of U2773-OT and SN~2009ip in the wavelength range
  of the [Ca~{\sc ii}] $\lambda\lambda$7291, 7325 lines
  (histograms).  The spectra are normalized, and SN~2009ip has a value
  of 0.5 subtracted for display.  The noisy gray spectrum is the HIRES
  spectrum of U2773-OT on day 22, in which the [Ca~{\sc ii}] lines fell
  near the edge of an echelle order.  The widths of the lines in
  U2773-OT are consistent with the 350 km s$^{-1}$ FWHM of H$\alpha$.}
\label{fig:ca2}
\end{figure}
%%%%%%%%%%%%%%%%%%%%%%%%%%%%%%%%%%%%%%%%%%%%%%%%%%%%%%%%%%%%%%%%%%%%%%%

The narrow H$\alpha$ and [N~{\sc ii}] features have Gaussian FWHM
$\approx$ 49 km s$^{-1}$.  The red [S~{\sc ii}] doublet is also
detected with the same line width (not shown).  The narrow lines
present in the HIRES extracted spectrum seem to indicate an underlying
H~{\sc ii} region, although 49 km s$^{-1}$ is quite broad for a simple
H~{\sc ii} region, so these lines may also arise in an extended
circumstellar nebula ejected by the star in a previous outburst.  This
narrow emission was also seen and spatially resolved in our LRIS
spectra, but the extended emission along the slit and the good seeing
allowed us to carefully subtract it.  Some [N~{\sc ii}] residuals
remain in Figure~\ref{fig:lrisHa}, but the narrow lines are
considerably weaker than in Figure~\ref{fig:hires1}.  In these narrow
lines, the observed [S~{\sc ii}] $F_{6716}/F_{6731}$ ratio is
$\sim$1.2, corresponding to a fairly low electron density of $\sim$120
cm$^{-3}$ (e.g., Osterbrock 1989).  The presence of an H~{\sc ii}
region coincident with U2773-OT supports the hypothesis that the
progenitor of U2773-OT was a young, massive star.  Both $\eta$ Car and
V1 in NGC~2363 are also located within bright, giant H~{\sc ii}
regions.  On the other hand, if this narrow emission arises in part
from extended circumstellar material, the [N~{\sc ii}] lines and other
features are reminiscent of extended LBV nebulae, which tend to be
enriched in N (Stahl 1987; Stahl 1989; Stahl \& Wolf 1986).

Finally, we show the spectrum in the region of the [Ca~{\sc ii}] lines
in Figure~\ref{fig:ca2}.  These lines were first noted to be strong by
Berger \& Foley (2009) on day 16.  Figure~\ref{fig:ca2} indicates that
the [Ca~{\sc ii}] lines are detected in both the HIRES spectrum (day
22) and the LRIS spectrum (day 34) as well.  The lines have width of
roughly 350 km s$^{-1}$, comparable to H$\alpha$, although the lines
are irregularly shaped, so accurate FWHM values are difficult to
determine.  Another unidentified emission line appears to be seen
between the two [Ca~{\sc ii}] lines.

\subsubsection{SN~2009\lowercase{ip}}

Figure~\ref{fig:lrisHa} shows the H$\alpha$ profile of SN~2009ip in
the medium-resolution LRIS spectrum, obtained on day 25 just after it
recovered from its sharp 3 mag dip.  This is superior to the HIRES
spectrum on 2009 Sep.\ 11 (Fig.~\ref{fig:hires2}), which had low S/N
because it was obtained during the sharp dip.  The H$\alpha$ line is
extremely bright, with a total emission equivalent width of
$-$198~\AA, and a flux of $2.1 \times 10^{-14}$ erg s$^{-1}$
cm$^{-2}$.  (This line flux is 3 times brighter than measured in the
HIRES spectrum.)  This corresponds to a total H$\alpha$ line
luminosity of $\sim2.5 \times 10^5$ L$_{\odot}$.  The H$\alpha$ line
profile is qualitatively similar several days earlier in the HIRES
spectrum obtained during the dip, but the line flux is weaker.

The H$\alpha$ line in SN~2009ip is clearly broader than that of
U2773-OT, and it is symmetric, unlike the P~Cygni absorption profile
seen in U2773-OT.  In order to fit the symmetric profile, one would
need a composite Gaussian with a broad, intermediate, and narrow
component.  However, the line profile is fit quite naturally with a
single Lorentzian profile with FWHM = 550 km s$^{-1}$.  The same
Lorentzian profile adequately accounts for H$\alpha$ in the HIRES
spectrum, although that spectrum has a much higher noise level.  As
discussed in detail by Smith et al.\ (2010) for SN~2006gy, a
Lorentzian profile is probably indicative of multiple electron
scattering through an opaque wind, suggesting a very high mass-loss
rate for the SN~2009ip eruption.  Although the H$\alpha$ line wings
extend to roughly $\pm$2000 km s$^{-1}$ at zero intensity, the
Lorentzian profile suggests that these are electron-scattering wings
and not true kinematic speeds of outflowing material.

We therefore adopt 550 km s$^{-1}$ as the characteristic outflow speed
of SN~2009ip, in agreement with the earlier assessment by Berger et
al.\ (2009b), although they did not specifically mention the
Lorentzian profile.  This speed is faster than the 300--370 km
s$^{-1}$ expansion observed in U2773-OT, V1 in NGC~2363 (Drissen et
al.\ 2001), and V37/SN~2002kg (Van Dyk et al.\ 2006), and it is closer
to the 600 km s$^{-1}$ outflow speed around more massive LBVs like the
Homunculus of $\eta$~Car (Smith 2006).  Despite this connection,
it is unclear if there is a trend of higher outflow speeds in more
luminous objects, because the outflow speeds were even higher in
SN~2008S and N300-OT, where the progenitors were less luminous.  In
Figure~\ref{fig:ca2} we also show the medium-resolution LRIS spectrum
in the wavelength range corresponding to the red [Ca~{\sc ii}] lines,
plotted along with the same wavelength range in U2773-OT.  It is clear
that the [Ca~{\sc ii}] lines are not detected in SN~2009ip.

%%%%%%%%%%%%%%%%%%%%%%%%%% FIGURE 12 - lowres spec Na I D  %%%%%%%%%%%%
\begin{figure}
\epsscale{0.95 }
%\plotone{../SPECTRA/nad.eps}
\plotone{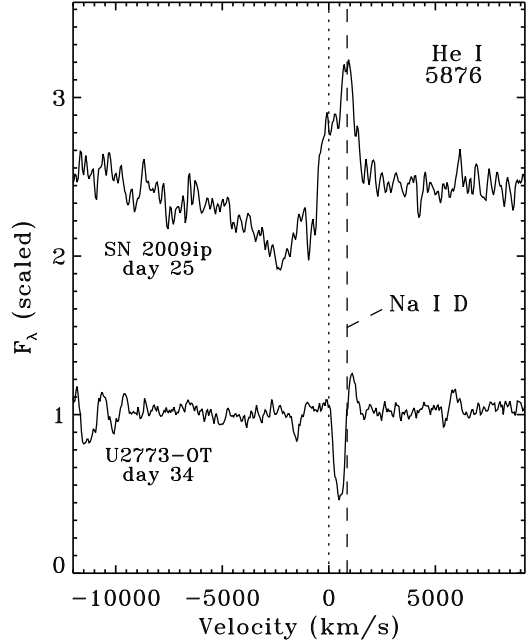}
\caption{Low-resolution LRIS spectra of SN~2009ip and U2773-OT, zooming
  in on He~{\sc i} $\lambda$5876 and the Na~{\sc i} D lines (the
  continuum level in the SN~2009ip spectra has been normalized with a
  tilted blue continuum). The velocity scale is relative to He~{\sc i}
  $\lambda$5876 (dotted line), but the zero-velocity position for
  Na~{\sc i} D $\lambda$5895 is shown with the long-dash line for
  reference.  The observed line profile is a combination of both lines,
  but we suspect that the broad emission and absorption feature in
  SN~2009ip is mainly He~{\sc i}, superposed with narrow Na~{\sc i} D
  emission and absorption.}
\label{fig:nad}
\end{figure}
%%%%%%%%%%%%%%%%%%%%%%%%%%%%%%%%%%%%%%%%%%%%%%%%%%%%%%%%%%%%%%%%%%%%%%%

\subsubsection{Fast Ejecta and a Blast Wave?}

Most spectral lines in both transients discussed here agree with the
outflow velocities determined from H$\alpha$, discussed above, and we
take these to be the dominant outflow speeds for each object.  In
U2773-OT the characteristic speeds are of order 350 km s$^{-1}$, while
the characteristic speeds of 550 km s$^{-1}$ for SN~2009ip are higher.
The H$\alpha$ profile in SN~2009ip does show wings extending to
$\pm$2000 km s$^{-1}$, but these seem consistent with
electron-scattering wings in a Lorentzian profile, as discussed above.
There are, however, signs of some faster material in SN~2009ip.

Figure~\ref{fig:nad} shows the region of the spectrum including
He~{\sc i} $\lambda$5876 and the Na~{\sc i} D lines, seen in the
low-resolution LRIS spectra of both targets (this spectral window was
not included in the blue or red medium-resolution LRIS spectra in
Figs.~\ref{fig:lrisHB} and \ref{fig:lrisHR}).  The Na~{\sc i} D line
in U2773-OT shows a narrow P Cygni profile, with an absorption trough
consistent with outflow speeds in other lines; there is no sign of
He~{\sc i} emission or fast material.

SN~2009ip, on the other hand, shows interesting new structure in this
line.  Aside from a narrow emission peak at the systemic velocity of
Na~{\sc i} D, the feature is dominated by a broad He~{\sc i}
$\lambda$5876 P~Cygni profile.  The emission component has a width and
wings consistent with other He~{\sc i} lines in the spectrum like
$\lambda$6678 and $\lambda$7065, as well as the symmetric wings of
H$\alpha$.  The blueshifted absorption of He~{\sc i} $\lambda$5876,
however, is the only line detected in our spectra that provides clear
evidence for faster outflow speeds of roughly 3000--5000 km s$^{-1}$.

These outflow speeds seen only in absorption exceed the characteristic
speed of 550 km s$^{-1}$ seen in emission lines in the spectrum of
SN~2009ip. This is reminiscent of the two ranges of outflow speeds
seen in $\eta$ Car: most of the mass in the Homunculus nebula expands
at 500--600 km s$^{-1}$ (Smith 2006), whereas recent spectra of faint
material exterior to that reveal much faster material moving at
3000--6000 km s$^{-1}$ (Smith 2008).  The kinematics are consistent
with both components originating in the same event in the 1840s,
implying that $\eta$ Car's giant eruption also had a fast blast wave
containing a comparable amount of kinetic energy but far less mass
than the slower Homunculus nebula (Smith 2008).  The fast 3000--5000
km s$^{-1}$ material seen in absorption in SN~2009ip, along with the
dominant speeds of 550 km s$^{-1}$ in most emission lines, may suggest
a similar scenario with a blast wave ahead of the slower ejecta for
SN~2009ip.  Indeed, the coexistence of narrow components of 550 km
s$^{-1}$ along with an intermediate-width component of a few 10$^3$ km
s$^{-1}$ is also reminiscent of the broader class of SNe~IIn (see the
discussion of line profiles in Smith et al.\ 2010 and Chugai \&
Danziger 1994), although the intermediate-width components tend to be
stronger in SNe~IIn.

If this fast material is evidence for a weak blast wave in SN~2009ip,
then there are several interesting implications beyond the connection
to $\eta$~Car.  First, shocks can heat material above the
equilibrium radiation temperature and may produce additional ionizing
radiation, so this might help explain the duality of spectra seen in
U2773-OT and SN~2009ip.  Namely, it may help explain why SN~2009ip had
a relatively hot spectrum, even though super-Eddington LBV eruptions
are expected to appear cooler (more like U2773-OT; see Humphreys \&
Davidson 1994).  Second, the existence of a shock on day 25 (during
the second peak) may hint that a shock-breakout event could be
responsible for the initial rise to peak luminosity and astonishingly
sudden decline, whereas the resurgence after the dip might then be
explained by the beginning of shock interaction with surrounding
circumstellar material, as in standard SNe~IIn.  A shock breakout has
never before been claimed in an LBV eruptive event, and the suggestion
is still quite speculative, but it may nevertheless have some
application to other events.  Unfortunately, we are not able to obtain
spectra or multi-band photometry during the initial peak of SN~2009ip,
so we cannot determine if the temperature was very high as one might
expect from shock breakout, but SN~2009ip warns that we should be on
the lookout for this behavior in future LBV eruptions.

%%%%%%%%%%%%%%%%%%%%%%%%%% FIGURE 13 - lowres spec %%%%%%%%%%%%%%%%%%%%%%%%
\begin{figure*}
\epsscale{1.1 }
%\plotone{../SPECTRA/spec09ip.eps}
\plotone{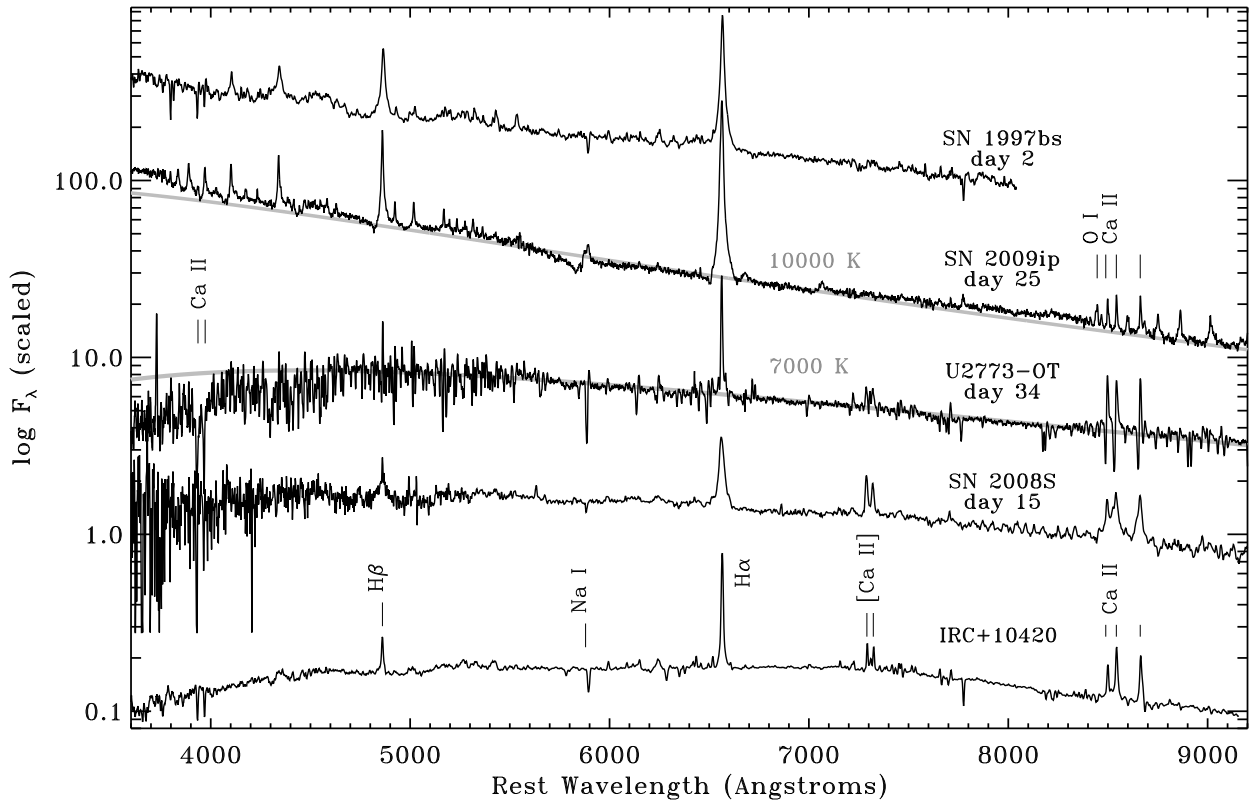}
\caption{Low-resolution LRIS spectra of SN~2009ip and U2773-OT,
  dereddened by $E(B-V)$ values of 0.019 mag and 0.564 mag,
  respectively. The continuum slopes of SN~2009ip and U2773-OT are
  matched by blackbodies with $T$ = 10,000~K and 7000~K, respectively.
  These spectra are compared to optical spectra of SN~2008S and the
  yellow hypergiant IRC+10420 (from Smith et al.\ 2009), as well as
  SN~1997bs (Van Dyk et al.\ 2002).}
\label{fig:spec}
\end{figure*}
%%%%%%%%%%%%%%%%%%%%%%%%%%%%%%%%%%%%%%%%%%%%%%%%%%%%%%%%%%%%%%%%%%%%%%%

\subsection{{\rm [Ca~{\sc ii}]} and IR Excess --- or Not}

Our low-resolution spectra of SN~2009ip and U2773-OT are shown in
Figure~\ref{fig:spec}.  These were obtained on the same night with
Keck/LRIS, and correspond to days 25 and 34 after discovery for
SN~2009ip and U2773-OT, respectively.  SN~2009ip is dominated by a
very blue ($T \approx$ 10,000~K) continuum and bright Balmer emission
lines, characteristic of LBVs.  It is compared to the spectrum of the
LBV eruption SN~1997bs in Figure~\ref{fig:spec}.  The dereddened
low-resolution spectrum of U2773-OT shows a blue continuum as well,
with a lower apparent temperature around 7000~K.  It also has Balmer
emission lines, although much narrower as noted earlier and riddled
with many deep blueshifted absorption features indicating an
underlying spectrum like an F supergiant with strong wind lines,
characteristic of many LBVs (see \S 3.3), as well as emission from
[Ca~{\sc ii}] and the Ca~{\sc ii} near-IR triplet.  The Ca~{\sc ii} H
\& K lines show strong, broad absorption, like SN~2008S and N300-OT.
The spectrum of U2773-OT in Figure~\ref{fig:spec} is similar to those
of SN~2008S and IRC+10420 (and by extension, N300-OT), although with
somewhat weaker [Ca~{\sc ii}] lines, and with more pronounced and more
numerous P~Cygni absorption features.

Our spectra of SN~2009ip and U2773-OT agree qualitatively with
preliminary reports of the spectra.  Berger et al.\ (2009b) obtained a
medium-resolution spectrum of SN~2009ip shortly after discovery, and
noted that it exhibited narrow (FWHM $\approx$ 550 km s$^{-1}$) Balmer
emission lines, in agreement with our spectrum on day 25.  Berger \&
Foley (2009) obtained spectra of U2773-OT and noted similarly narrow
(FWHM $\approx$ 350 km s$^{-1}$) Balmer emission lines, in agreement
with our HIRES and LRIS spectra described above, plus [Ca~{\sc ii}],
Ca~{\sc ii}, and a few other emission features.  Taken at face value,
one might conclude from these spectral characteristics that U2773-OT
is closely related to the recent transients SN~2008S and N300-OT,
which were discovered to have unusually bright [Ca~{\sc ii}] emission
in their spectra (Smith et al.\ 2009; Berger et al.\ 2009a; Bond et
al.\ 2009), whereas SN~2009ip is apparently not.

The other very unusual characteristic shared by SN~2008S and N300-OT
is that both had heavily dust-enshrouded progenitors detected only in
{\it Spitzer} data (Prieto 2008; Prieto et al.\ 2008), and significant
dust was still present around both at late times (Prieto et al.\ 2009;
Wesson et al.\ 2010). Does this connection between [Ca~{\sc ii}]
emission and dust hold for the two new transients discussed here?

SN~2009ip was detected at $\sim$20.2 mag in an unfiltered Keck guider
image on 2009 Sep.\ 11. Assuming this guider image closely
approximates the $R$ band, this means the $R-K'$ color was $\sim$0.5
mag at this epoch, compared to our near-IR image taken 2 days earlier.
Given the fast optical decline of SN~2009ip at this time ($\sim$2 mag
between Aug.\ 30 and Sep.\ 11), however, it is likely that $R-K$ was
even less than 0.5~mag. Thus, we measure very little if any near-IR
excess in SN~2009ip.  It must have had little circumstellar dust near
the star to cause excess emission or circumstellar extinction, in
direct contrast to N300-OT and SN~2008S which had $R-K \approx 3$ mag
and $\sim$1.7 mag, respectively, at peak optical output (Bond et
al.\ 2009; Botticella et al.\ 2009). For both of these transients the
$R-K$ color evolved redward as they declined.

U2773-OT, on the other hand, does show a considerable near-IR excess.
After correcting for Galactic extinction, we find that U2773-OT had
$R-K_s \approx 1.5$ mag on 2009 Sep.\ 8.  Figure~\ref{fig:ir}
shows that its near-IR/optical color has remained relatively unchanged
since then, as it brightened by $\sim$0.2 mag over 100 days.  Its
near-IR excess is comparable to that of SN~2008S in eruption.
Regardless of whether the near-IR color excess is caused by reddening
from circumstellar dust or emission from heated dust, the IR excess
points to a dusty environment around U2773-OT. (The IR excess might
also be caused in part by newly formed dust, although a more detailed
study of the evolving spectral energy distribution and line profiles
would be needed to investigate this further.) 

It therefore appears that U2773-OT did indeed have a substantially
dusty circumstellar environment to accompany its [Ca~{\sc ii}]
emission --- like SN~2008S and N300-OT --- while SN~2009ip had
neither.  The dusty environment around the U2773-OT progenitor was not
as fully obscuring as the dust around the progenitors of SN~2008S and
N300-OT, although in those cases the progenitor obscuration was due
mostly to a more compact distribution of dust that seems to have been
largely destroyed by their luminous outbursts (Wesson et al.\ 2010;
Prieto et al.\ 2009), leaving more distant dust shells comparable to
that seen now around U2773-OT.  The dust around U2773-OT is also
reminiscent of the dusty environments inferred around the transients
M85-OT and SN~1999bm, which led Thompson et al.\ (2009) to place those
two objects in the same class with SN~2008S and N300-OT.

This is intriguing, since we have shown that {\it both} SN~2009ip and
U2773-OT were otherwise very similar, with pre-eruption LBV
variability in the decade before discovery, luminous LBV-like
progenitors, and very narrow line widths that were even narrower than
those of SN~2008S or N300-OT but characteristic of LBVs.  One must
therefore conclude that the presence of strong [Ca~{\sc ii}] and
Ca~{\sc ii} emission {\it cannot be taken as evidence for or against
  an LBV nature}.  Instead, [Ca~{\sc ii}] emission is more closely
related to the progenitor's circumstellar environment --- especially
the presence of dust.  It was already noted that the bright [Ca~{\sc
  ii}] and Ca~{\sc ii} lines in SN~2008 and N300-OT are probably
related to dense circumstellar gas and pre-existing circumstellar dust
grains that were vaporized by the increased outburst radiation (Prieto
et al.\ 2008; Smith et al.\ 2009).  The corresponding implications for
SN~2008S and N300-OT are discussed further in \S 4.2.

A dusty circumstellar environment around the progenitor of U2773-OT
affects our estimates of the progenitor properties, as noted above.
From {\it HST} photometry corrected only for Galactic extinction and
reddening, we found that the quiescent progenitor ($\sim$10 yr before
discovery) had a color consistent with that of an early A-type
supergiant having a luminosity of log$(L/{\rm L}_{\odot}) \approx
5.1$, and an initial mass of roughly 20~M$_{\odot}$.  With substantial
circumstellar dust, the progenitor of U2773-OT must have been even
hotter than an early A-type star, and more luminous and massive as
well, although one cannot make precise corrections with available
information.  It is difficult to escape the conclusion that U2773-OT
was blue, quite luminous, and obviously highly variable, so it was
therefore likely to have been an LBV.  Its spectral morphology
confirms this.  Conversely, there was little pre-existing
circumstellar dust around SN~2009ip, which is reassuring because its
quiescent progenitor with $M \approx -10$~mag would already be among
the most luminous stars known, and any significant amount of dust
would raise the corresponding progenitor luminosity.

%%%%%%%%%%%%%%%%%%%%%%%%%%%%%%%%%%%%%%%%%%%%%%%%%%%%%%%%%%%%%%%%%%%%%%%%%
\section{DISCUSSION}

\subsection{Precursor S~Doradus Eruptions}

A critical new result from this study is that we have recovered the
luminous progenitor stars and precursor variability in the decade
preceding the discovery of SN~2009ip and U2773-OT, and
that the luminosity and variability are characteristic of known LBVs
like $\eta$~Car, SN~1954J (V12), and V1 in NGC~2363.  Before these
precursor outbursts began, the two stars were apparently at quiescent
absolute magnitudes of $-8.0$ to $-10$ mag, implying moderately massive
and very massive progenitor stars for U2773-OT and SN~2009ip,
respectively.

We propose that the apparent brightening episodes in the $\sim$10 yr
before the giant LBV eruptions of SN~2009ip and U2773-OT were caused by
preparatory outburst phases, akin to S~Dor-like variability, which
then grew into giant eruptions as is thought to be the case for the
precursor behavior of $\eta$~Car and SN~1954J (Humphreys et al.\
1999). SN~1961V may have also been in an S~Dor phase for 10--20 yr
preceding its giant eruption (Goodrich et al.\ 1989).

The visual brightening identified as S~Dor outbursts, named for
the famous prototype LBV star in the Large Magellanic Cloud, occurs
when the star varies at roughly constant $M_{\rm Bol}$, but brightens
roughly 1--2.5 mag at visual wavelengths because it becomes cooler,
redistributing its peak flux from the UV to visual wavelengths (see
Humphreys \& Davidson 1994).  S~Dor outbursts are a {\it defining}
observed phenomenon in LBVs, although not every S~Dor phase is
followed by a giant eruption.  The namesake of the class, for example,
has never been observed in a giant eruption, but only smaller
oscillations.  The situation can be more complicated as well: 
S~Dor-like variability may evolve into larger outbursts.  For example, 
the 1990s eruption of V1 in NGC~2363 may not have been a simple S~Dor
phase, because its bolometric luminosity increased (Drissen et al.\
2001), and we should be mindful that this may be the case for our two
new transients discussed here as well.  In any case, LBVs exhibit a
wide variety in $\Delta M_V$ and $\Delta t$, but a few to 10 yr is a
typical observed timescale for these S~Dor variations (see Humphreys
\& Davidson 1994; van Genderen 2001).

As noted earlier, neither S~Dor outbursts nor giant LBV eruptions have
a theoretical explanation, although they are suspected to be the
result of luminous stars flirting with an opacity-modified Eddington
limit and subsequent runaway mass loss (e.g., Smith \& Owocki 2006;
Humphreys \& Davidson 1994; Smith et al.\ 2003; Owocki et al.\ 2004).
SN~2009ip and U2773-OT are important in this context because they add
two new well-observed cases to the three historical examples of
precursor variability before a giant LBV eruption, and unlike the
historical examples, we also have obtained spectra of their eruptions
which resemble other recent SN impostors without documented
pre-eruption LBV-like variability.

Given the stark differences from one object to the next, and the
presence of ``preparatory'' precursor outbursts in the decade
preceding maximum light for some objects, one wonders to what extent
there is cumulative hysteresis built into the system.  For example, to
what extent does the peak luminosity output and rate of decline depend
systematically on the pre-eruption variability and its associated mass
loss?  At this time such questions are still quite speculative, but a
larger sample of these transients with good observations would
obviously be valuable to establish any such trends.

\subsection{Implications for SN~2008S and N300-OT}

The spectrum of one object we discussed here, U2773-OT, resembles the
optical spectra of SN~2008S, N300-OT, and IRC+10420, as noted above.
The other, SN~2009ip, matches spectra of SN impostor LBVs like
SN~1997bs that are dominated by Balmer lines alone.  Yet, based on the
luminous progenitors and their S~Dor variability, we have identified
{\it both} SN~2009ip and U2773-OT as giant LBV eruptions.

This link reinforces earlier suggestions that SN~2008S and N300-OT may
also have been related to LBV-like eruptions, albeit with relatively
low-luminosity (and possibly cooler) dust-obscured progenitors.  It
also suggests that some giant LBV eruptions have bright [Ca~{\sc ii}]
and Ca~{\sc ii} emission lines while some do not.  Several authors
have noted that the bright [Ca~{\sc ii}] and Ca~{\sc ii} lines are
probably related to dense circumstellar gas and the destruction of
dust grains (Prieto et al.\ 2008; Smith et al.\ 2009; this work).  In
the case of IRC+10420, this dust destruction\footnote{Botticella et
  al.\ (2009) stated that IRC+10420 is not enshrouded by a dust shell
  to support their claim that it is not related to SN~2008S and
  N300-OT.  In fact, IRC+10420 is heavily enshrouded by circumstellar
  dust, with most of its luminosity escaping at 10--20 $\mu$m (e.g.,
  Humphreys et al.\ 1997). Although IRC+10420 is more luminous, the
  ratio between its 8 $\micron$ and optical $R$-band flux is similar
  to the observed colors of SN~2008S and N300-OT.}  occurs because the
star has been growing much warmer over the past 30 years, and
increasing its UV output (see Smith et al.\ 2009; Humphreys et al.\
2002).  In the transients, it occurs because of the sudden increase in
luminosity that pushes the grain sublimation radius farther into the
progenitor's dust shell.

Among known LBVs, some are dusty, and some are not.  Some have no
detected circumstellar shells, while others like $\eta$~Car have 15
M$_{\odot}$ dust shells (Smith et al.\ 2003) that reprocess nearly all
of the star's UV and visual radiation.  The difference depends on the
star's previous mass-loss history (i.e., how recently it has ejected a
massive dust shell, and how fast that dust shell expands and thins).
Regardless of its initial mass, a star that has recently suffered an
LBV-like eruptive event could be completely obscured at optical
wavelengths, while its IR/optical colors would evolve rapidly over the
subsequent century.  The obscured phase could be very brief and mostly
missed in samples of known LBVs, in agreement with the expected short
duration implied by the rarity of similar dust obscured stars
(Thompson et al.\ 2009).  If strong [Ca~{\sc ii}] emission is indeed
linked to vaporizing circumstellar dust, then it would not be
surprising to have some giant LBV eruptions with strong [Ca~{\sc ii}]
emission and some with only weak emission or none.  For example,
Valeev et al.\ (2009) recently reported a new LBV in M33, which has a
dust shell and bright [Ca~{\sc ii}] emission.

Another key issue is the variability of the progenitors.  The lack of
IR variability in the progenitors of SN~2008S and N300-OT seems
problematic for their interpretation as EAGB stars that become ecSNe.
All stars at the tip of the AGB pulsate with large amplitudes (e.g.,
Mira variables).  Thompson et al.\ (2009) demonstrated clearly that
over timescales of a few years in the same IRAC bands, known EAGB
stars are highly variable ($\Delta M_{4.5}$ is typically $\sim$1 mag),
whereas LBV candidates with dust shells are not highly variable (a few
exceptions have $\Delta M_{4.5}$ as much as 0.5 mag).  One reason for
this is that although LBVs are variable at visual wavelengths during
S~Dor-like episodes, their {\it bolometric} luminosity remains roughly
constant.  Recall that the observed visual-wavelength brightening in
S~Dor outbursts is thought to be a redistribution of flux from the UV
to optical --- i.e., a change in bolometric correction at constant
luminosity.  Since the dust properties depend primarily on the central
engine's bolometric output, one may expect the mid-IR luminosity of a
circumstellar dust shell around an LBV to be nearly constant --- that
is, until it experiences a giant eruption when the bolometric
luminosity actually climbs.

SN~1954J/V12 is a key example to keep in mind when debating the nature
of SN~2008S and N300-OT, because it is a well-established LBV, but its
luminosity was comparable to the relatively low luminosities of the
two dust-enshrouded transients.  The two new sources we report here
appear to bridge the gap between these examples and more luminous
LBVs.  If SN~2008S and N300-OT were blue supergiants that were heavily
obscured by dust, then the initial masses implied by their IR
luminosities are higher than if one assumes that they are at the tip
of an ascending AGB/RSG branch.  One cannot confidently determine the
effective temperature or color of an underlying star from the mid-IR
dust emission properties; for example, in the mid-IR color plots
presented by Thompson et al.\ (2009), $\eta$ Car had the same {\it
  Spitzer} colors as the very cool red supergiant VY~CMa.  Altogether,
then, we find the case to be still quite plausible that SN~2008S and
N300-OT may have been eruptions analogous to episodic LBV-like events,
at lower initial masses than had previously been recognized.  The
variety among this class of objects may be telling us that episodic
mass loss and eruptive phenomena are a generic property of late
evolutionary phases over a wide range of initial masses, whatever the
underlying cause.

Still, there are some key differences between LBVs and
SN~2008S/N300-OT which remain unsolved even if these are not episodic
LBV-like eruptions.  One is the apparently C-rich chemistry in the
circumstellar dusty envelope of N300-OT (Prieto et al.\ 2009).
Non-silicate dust was also inferred for SN~2008S (Wesson et al.\
2010).  Massive stars are generally not found to be C-rich for long
periods except in the WC Wolf-Rayet phase, providing the most
compelling argument made for a connection to EAGB stars (Prieto et
al.\ 2009; Thompson et al.\ 2009).  This is puzzling as well, though,
since C-rich AGB stars are thought to come from lower-mass stars (Jura
1991), well below the N300-OT progenitor mass of 12--25 M$_{\odot}$
inferred by Gogarten et al.\ (2009).  Furthermore, the C enrichment in
single carbon stars is thought to arise from dredge up during unstable
He shell burning over a degenerate C-O core.  One expects progenitors
of ecSNe to avoid this enrichment, however, because they do not make a
degenerate C-O core, instead burning carbon to eventually produce the
O-Ne-Mg core that collapses (e.g., Nomoto 1984).  With such mysteries,
one may feel tempted to appeal to previous episodes of mass transfer
in evolved close binary systems for any plausible initial mass range.
It is worth noting, however, that observations of the dust composition
in $\eta$ Car (Chesneau et al.\ 2005; Mitchell \& Robinson 1978; Smith
2010) have also shown strong evidence for unusual non-silicate dust,
such as corundum (Al$_2$O$_3$) and other species.  The unusual grain
composition may be related to the {\it rapid} formation of dust around
a hot star, where condensation temperature may compete with chemical
abundances in determining the grain composition (e.g., Smith 2010).
In other words, the presence of carbon dust may not necessarily result
directly from carbon-rich gas-phase abundances.  For many other LBVs,
the dust composition has not been studied in detail yet, so an attempt
to detect or rule out the presence of carbon-bearing dust or molecules
would be interesting.

\section{CONCLUSIONS}

We have investigated the two recent transients SN~2009ip and U2773-OT.
While they show some spectral differences, we conclude that they are
two manifestations of the same underlying phenomenon: namely, they
are both giant eruptions of luminous blue variables (LBVs).  Here we
briefly summarize the main conclusions of our study.

(1) The quiescent progenitor stars of both transients have been
identified in {\it HST} images taken $\sim$10~yr before discovery.
The progenitor of SN~2009ip was an extremely luminous star with $M_V
\approx -9.8$~mag, log($L/{\rm L}_{\odot}) = 5.9$, and a probable
initial mass of 50--80 M$_{\odot}$.  This places it well above the
observed upper limit for red supergiants, so it was either a yellow or
blue supergiant.  The progenitor of U2773-OT was somewhat less
luminous, at $M_V \approx -7.8$~mag and log($L/{\rm L}_{\odot}) =
5.1$, corresponding to an initial mass of $\sim$20 M$_{\odot}$.  It
had an observed color similar to that of an early A-type supergiant,
consistent with lower-luminosity LBVs.  However, an IR excess reveals
that U2773-OT had a dusty environment, so it was most likely hotter,
more luminous, and more massive, also consistent with an LBV.

(2)  Examining pre-discovery ground-based data, we discovered a long
$\sim$5~yr history of LBV-like photometric variability in the decade
leading up to the peaks of eruption for both transients.  The
precursor variability resembles the preparatory S~Dor phases that
transformed into giant LBV eruptions of $\eta$~Car, SN~1954J, and
SN~1961V.  We suspect that this early warning sign is an important key
for triggering a giant eruption, although many S~Dor phases do not
result in a giant eruption, and it is unclear if all giant eruptions
show precursor outbursts.

(3) Immediately after reaching its peak, we discovered that SN~2009ip
took a sharp dive, fading by 3.2 mag in 16 days, followed by an almost
immediate recovery to nearly its previous level.  The astonishing
decline rate is faster than that of SNe, and the recovery is
unprecedented.  The closest analog is the visual light curve of $\eta$
Car during its 1843 Great Eruption.  This would suggest that
SN~2009ip is physically related to $\eta$ Car, and we should not
be surprised if its eruption continues.  After reaching its peak
level, U2773-OT seems to have sustained a relative plateau, with only
minor variations at the time of writing.

(4)  Optical spectra of the two transients are qualitatively different, 
but each is characteristic of LBVs at different phases of their
variability. SN~2009ip is dominated by very strong Balmer emission
lines and outflow speeds of $\sim$550 km s$^{-1}$, similar to $\eta$
Car. Evidently, SN~2009ip did not show a cool photosphere, counter
to some expectations for LBV eruptions.  U2773-OT has a cooler
7000--8000~K spectrum of an F supergiant superposed with wind emission
lines and sharp blueshifted absorption features, indicating an
outflowing wind of 350 km s$^{-1}$.  It is almost identical to the
spectrum of S~Dor and other LBVs in their cool eruptive phases.  This
dichotomy illustrates the wide variety of spectral properties that can
be observed in LBV eruptions.

(5) SN~2009ip also shows evidence for fast material moving at
3000--5000 km s$^{-1}$ seen in absorption of He~{\sc i} $\lambda$5876.
These speeds are much faster than most of the mass traced by emission
lines (550 km s$^{-1}$), and this combination is quite reminiscent of
the slower Homunculus and fast blast wave of $\eta$ Car (Smith 2008).
We speculate that shock excitation by this fast blast wave may play a
role in giving rise to the hotter spectrum of SN~2009ip, and may be
related to the sharp initial peak in the light curve.

(6) The spectrum of U2773-OT has [Ca~{\sc ii}] emission and other
features reminiscent of the recent transients SN~2008S and N300-OT, as
well as an IR excess indicating a dusty environment.  SN~2009ip has
neither of these attributes, yet both are clearly LBVs. This
illustrates that the diversity of LBV circumstellar environments can
give rise to a wide range of properties such as those seen in SN~2008S
and N300-OT, depending on the star's recent mass-loss history.

(7) We have noted that the progenitor luminosities for some LBVs are
much lower than that of $\eta$ Car, such as U2773-OT, SN~1954J/V12,
and V1 in NGC~2363, all of which imply initial masses around
20~M$_{\odot}$.  The progenitor luminosities of these well-established
LBVs overlap with the low IR luminosities of the dust-obscured
progenitors of SN~2008S and N300-OT, adding weight to the possibility
that these are moderately massive stars of 12--15 M$_{\odot}$ that
experienced eruptive instability analogous to LBVs.

%\medskip
\acknowledgments 
\footnotesize
%\scriptsize

We thank an anonymous referee for critical comments that helped
improve the manuscript.  Some of the data presented herein were
obtained at the W. M. Keck Observatory, which is operated as a
scientific partnership among the California Institute of Technology,
the University of California, and NASA; the observatory was made
possible by the generous financial support of the W. M. Keck
Foundation.  We thank the Lick and Keck Observatory staffs for their
dedicated help, and we acknowledge Chris Fassnacht for assistance with
obtaining flatfields for the NIRC2 observations.  We are grateful to
Cullen Blake, Dan Starr, and Mike Strutskie for their help with the
development and operations of PAIRITEL. In addition, we appreciate the
assistance and patience of Ryan Foley, Armin Rest, and Dan Kasen
during the Keck/LRIS observing run when we discovered the
rebrightening of SN~2009ip.  The supernova research of A.V.F.'s group
at U.C. Berkeley is supported by National Science Foundation (NSF)
grants AST-0607485 and AST-0908886, the TABASGO Foundation, and the
Richard \& Rhoda Goldman Fund.  Financial support for this work was
also provided by NASA through grant AR-11248 from the Space Telescope
Science Institute, which is operated by Associated Universities for
Research in Astronomy, Inc., under NASA contract NAS 5-26555.  KAIT
and its ongoing operation were made possible by donations from Sun
Microsystems, Inc., the Hewlett-Packard Company, AutoScope
Corporation, Lick Observatory, the NSF, the University of California,
the Sylvia \& Jim Katzman Foundation, and the TABASGO Foundation.
J.S.B. and A.A.M. were partially supported by a NASA/{\it Swift} Guest
Investigator (Cycle 5) grant and a SciDAC grant from the
U.S. Department of Energy.  This research used data products from the
Two Micron All Sky Survey, which is a joint project of the University
of Massachusetts and the Infrared Processing and Analysis
Center/California Institute of Technology, funded by NASA and the
NSF. It also used resources of the National Energy Research Scientific
Computing Center, which is supported by the Office of Science of the
U.S. Department of Energy under Contract No.\ DE-AC02-05CH11231;
P.E.N. thanks them for a generous allocation of computing time and
space on their machines in support of DeepSky.

{\it Facilities:} {\it HST} (WFPC2), Keck I (LRIS, HIRES), Keck II
(NIRC2/LGSAO), Lick (KAIT), PAIRITEL

% REFERENCES

\end{document}